\documentclass[aps,a4paper,showpacs,amsfonts,amsmath,amssymb,nofootinbib]{revtex4}
\usepackage{makeidx}
\usepackage{graphicx}
\usepackage{epsfig}
\usepackage{xcolor}
\usepackage{placeins}
\usepackage{ulem}

\newcommand \beq {\begin{equation}}
\newcommand \eeq {\end{equation}}
\newcommand \bd {\begin{displaymath}}
\newcommand \ed {\end{displaymath}}
\newcommand \beqn {\begin{eqnarray}}
\newcommand \eeqn {\end{eqnarray}}
\newcommand \la {\langle}
\newcommand \ra {\rangle}

\begin{document}


\title{Counting of level crossings for inertial random processes:
Generalization of the Rice formula}

\author{Jaume Masoliver}
\email{jaume.masoliver@ub.edu}
\author {Matteo Palassini}
\email{palassini@ub.edu}
\affiliation{Department of Condensed Matter Physics and Institute of Complex Systems (UBICS), \\
University of Barcelona, Catalonia, Spain}

\date{\today}

\begin{abstract}

We address the counting of level crossings for inertial stochastic processes. We review Rice's approach to the problem and generalize the classical Rice formula to include all Gaussian processes in their most general form. We apply the results to some second-order (i.e., inertial) processes of physical interest, such as Brownian motion, random acceleration and noisy harmonic oscillators. For all models we obtain the exact crossing intensities and discuss their long- and short-time dependence. We illustrate these results with numerical simulations.

\end{abstract}


\maketitle

\section{Introduction}
\label{intro}

Level-crossing problems --and related issues such as hitting, extreme-value, first-passage and exit times problems, among others-- are not only of deep physical and theoretical interest, but also of considerable practical importance, with countless applications ranging from chemical physics, meteorology, seismology, reliability theory, structural and electrical engineering, and even economics and finance, just to name a few \cite{blake,redner_book,rychlik,mp_2014,maso_llibre,majumdar_pr_2020}. In a rather general form we may say that the level-crossing problem consists in gathering information on the interval between crossing points to some given level or mark --usually critical--  with the ultimate objective of obtaining the probability density of the time intervals between consecutive crossings, a problem which, unfortunately, has no known exact solution \cite{munakata}. What is however known (at least to some  extent) is the counting of level crossings.   

The problem of level-crossing counting was first thoroughly discussed during the mid nineteen forties by S. O. Rice \cite{rice,rainal} within statistical communication theory and it was restricted to stationary Gaussian processes. The main result was the classical Rice formula for the average number of occasions, per unit time, that these processes cross a given level. While Rice was primarily concerned with applications to electrical and radio engineering, the matter has deep and far-reaching effects on other fields of knowledge such as ocean and mechanical engineering, chemical physics, material sciences, laser physics and optics, and many more (see the review \cite{lindgren_2019}). After Rice the problem was  first put on firmer mathematical basis by It\^o \cite{ito}, Ylvisaker \cite{ylvisaker}, and particularly by the Scandinavian school of statistics led by Harald Cramer and collaborators \cite{rychlik,lindgren_2019,cramer,cramer_leadbetter,leadbetter_spaniolo,lindgren_book}, among others (see \cite{kratz,borovkov_08,borovkov_12} for a small sample).  

One of the main achievable goals in the theory of level crossings is provided by the crossing intensity, or average crossing frequency, which is the average number of times (per unit time)  that a random process crosses some given level. The inverse of such a quantity has dimensions of time and is called the return period. In mechanical engineering this is a key quantity since it measures the severity of the load on a given structure.  For instance, in ocean engineering, in designing walls for the protection against high sea levels the sea surface is generally modeled by stationary Gaussian fields with random excursions from an average height \cite{rychlik}.    

As we will recall in the next section, in order to develop Rice's approach to a given stochastic process, it is necessary to know the joint probability density of the process and its time derivative, which in many cases is not known. For example, first-order processes driven by white noise are not differentiable, thus this joint density does not exist. 
One of the objectives of this work is to extend Rice theory and obtain exact expressions of the crossing intensity for linear second-order (i.e., inertial) random processes.  

As far as we know, most applications and generalizations of Rice theory are restricted to Gaussian processes and extensions thereof. This is for instance the case of the Slepian model for Gaussian and stationary processes after crossings of the average level \cite{lindgren_slepian}. Another extension is addressed to quadratic sums of, again, Gaussian processes (the so-called $\chi^2$ processes \cite{rychlik}), which are important in modeling the response of a given structure to a wind load. In both extensions, solutions are usually numerical and essentially focused on engineering applications. Rice's formula can also be derived from the Kac counting formula \cite{kac} for the roots of functions with continuous first derivative, and for this reason it is sometimes called Kac-Rice formula \cite{azais,berzin}. The Kac formula has been generalized to scalar-valued random fields \cite{longuet} and vector-valued random fields \cite{azais,berzin}.

Rice's theory has been widely studied in mathematics and engineering but, to our knowledge, it seems to be less known in physics. Our main goals here are to review the theory using simple arguments and, as mentioned above, to apply it to inertial random process which naturally arise in many physical applications.
Previous physical applications of Rice's theory 
include persistence and first-passage properties (see review \cite{bray_etal} and references therein). The number of crossings of the order parameter at a given level has been used to analyze metastable states in the stochastic evolution of spin systems 
\cite{paul1, paul2}, but in this case the evolution is not inertial. Rice's theory was also generalized to determine the number of critical points in stochastic processes and random fields, such as those arising in the statistical physics of disordered systems \cite{fyodorov, fyodorov2, bray}.

The paper is organized as follows. In Sect. \ref{sec_2} we review the classical Rice formula of the crossing intensity. In Sect. \ref{sec_3} we obtain the most general expression of the crossing intensity for any Gaussian process. In Sect. \ref{sec_4} we apply the results to some particular but relevant Gaussian inertial processes such as Brownian motion and random acceleration process. Sect. \ref{sec_5} is devoted to random oscillators either damped and undamped with a thorough discussion on different time scales. Concluding remarks are in Sect. \ref{sec_6} and some technical details in three appendices.

\section{The level-crossing problem and Rice formula}
\label{sec_2}

Historically, the level-crossing problem stemmed from Rice zero-crossing problem \cite{rice,rainal} which in turn originated in Kac's search of the zeros of random polynomials \cite{kac}. Rice studied the case in which the random process was given by the explicit form $X(t)=f(a_1,\cdots,a_n;t)$ where $f(\cdot)$ is any given function and $a_1,\cdots,a_n$ are random variables. He then obtained an explicit expression for the average number of zeros per unit time when $X(t)$ is a stationary Gaussian process. 
The result was latter extended to wider classes of random processes, including non-stationary ones \cite{lindgren_book}.
We will next review the general formula for the counting of level crossings using intuitive arguments rather than a more rigorous mathematical reasoning. We essentially follow Rice original approach \cite{rice} as well as  Blake and Lindsey excellent review \cite{blake}, and refer the interested reader to Lindgren's textbook \cite{lindgren_book} for more rigorous derivations.

\subsection{Level-crossing intensity}

Let $X(t)$ be a random process and denote by $Y(t)=\dot X(t)$ its time derivative (also called velocity) which is supposed to exist, at least in the sense of generalized functions, and let $p(x,y,t)$ be the joint probability density function (PDF) of $X(t)$ and $Y(t)$. In a first step, the level-crossing problem consists in counting the number of times that $X(t)$ attains a certain level or mark $u$ (which can be time dependent), that is to say, in obtaining statistical information on the random quantity:
$$
N_u(t_0,t)=\mbox{ no. of times } X(\tau)=u, \quad (t_0\leq\tau\leq t).
$$
In some applications it is important to distinguish whether the crossing of level $u$ occurred while ``going up" or ``going down'', we thus have the number of {\it upcrossings},
$$
N_u^{(+)}(t_0,t)=\mbox{ no. of times } X(\tau)=u,\  \dot X(\tau)>0, \quad (t_0\leq\tau\leq t),
$$
and we can analogously define the number of {\it downcrossings} $N_u^{(-)}(t_0,t)$ in which $\dot X(\tau)<0$. These quantities are obviously random variables depending on the particular realization of the process $X(t)$. 

\begin{figure}[ht]
\includegraphics[width=0.6\linewidth,angle=0]{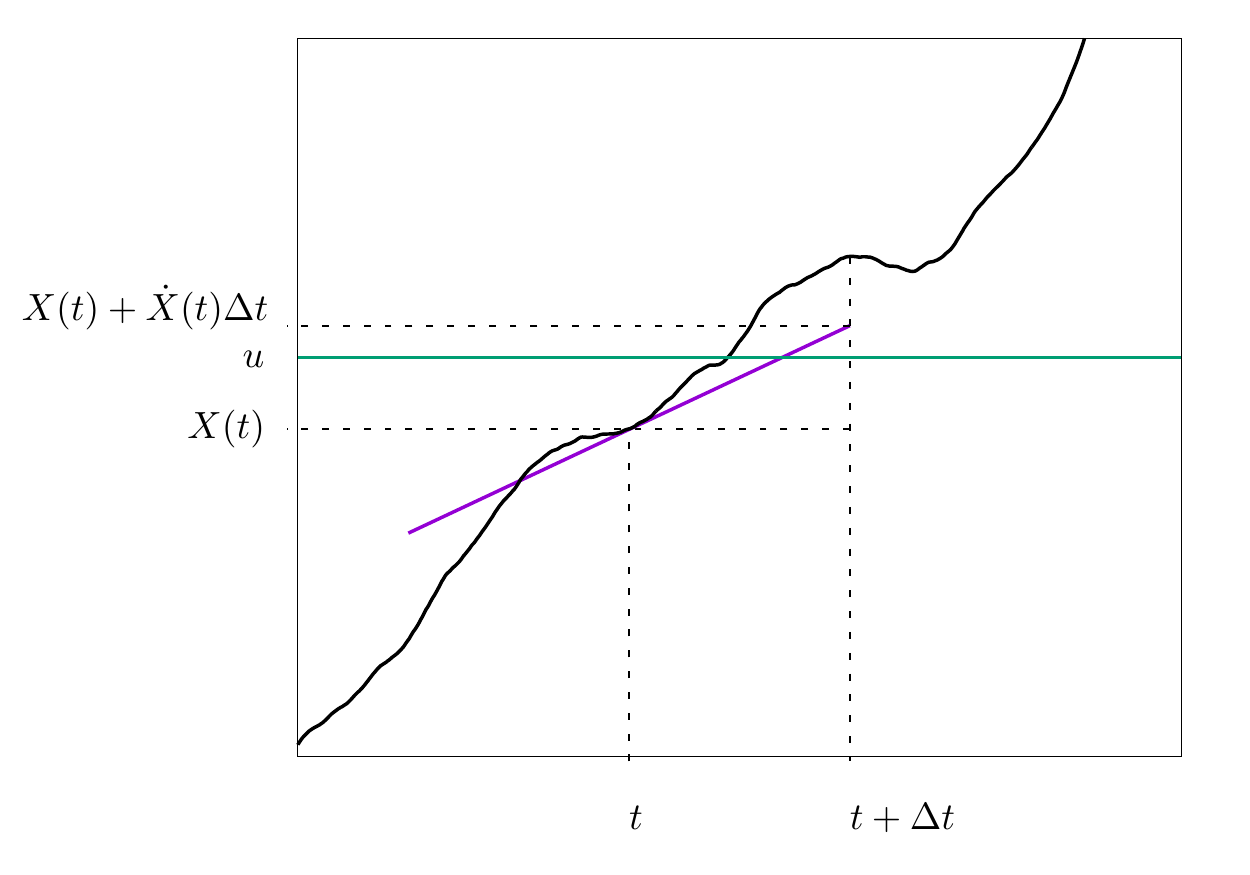}
\caption{Illustration of an upcrossing event. The irregular (black) line represents a simulated random trajectory $X(t)$, the straight  oblique (purple) line has slope $\dot{X}(t)$. If $\Delta t$ is small enough, $X(t)$ will cross the level $u$, represented by the horizontal solid (green) line in the interval $(t,t+\Delta t)$ if $\dot{X}(t)>0$ and  $u-\dot{X}(t)\Delta t\leq X(t)\leq u$.}
\label{drawing}
\end{figure}

We will now obtain the probability of having a crossing event to any level $u$ during a time interval $(t,t+\Delta t)$. Let us first observe that the probability of having more than one crossing during the interval is negligible as long as $\Delta t$ is small. Therefore, during small time intervals, the probability of having a crossing event equals the probability that $N_u(t,t+\Delta t)=1$. Let us also note that the crossing of any level $u$ for the process $X(t)$ during a small time interval $(t,t+\Delta t)$, will take place either (i) if $X(t)$ is between the positions $u-Y(t) \Delta t$ and $u$ while the velocity $Y(t)$ is positive (upcrossing), as illustrated in Fig. \ref{drawing}, or (ii) if $X(t)$ is between $u$ and $u+|Y(t)|\Delta t$ while $Y(t)$ is negative (downcrossing).

Consequently, the probability of a crossing during 
$(t,t+\Delta t)$, either down or up,  is
\begin{eqnarray*}
&&{\rm Prob}\Bigl\{ N_u(t,t+\Delta t)=1 \Bigr\} \\
&&={\rm Prob}\Bigl\{ u-Y(t)\Delta t\leq X(t)\leq u, Y(t)>0\Bigr\}+{\rm Prob}\Bigl\{ u \leq X(t)\leq u+|Y(t)|\Delta t, Y(t)<0\Bigr\} 
\end{eqnarray*}
or, in terms of the joint PDF $p(x,y,t)$,
\begin{eqnarray*}
{\rm Prob}\Bigl\{ N_u(t,t+\Delta t)=1 \Bigr\} &=&
\int_0^\infty dy\int_{u-y\Delta t}^u p(x,y,t) dx+\int_{-\infty}^0 dy\int_u^{u+|y|\Delta t} p(x,y,t) dx \\
&=& \Delta t\left[\int_0^\infty y p(u,y,t)dy
+\int_{-\infty}^0 |y| p(u,y,t)dy\right] + O(\Delta t^2), 
\end{eqnarray*}
that is,
\begin{equation}                %
{\rm Prob} \Bigl\{ N_u(t,t+\Delta t)=1 \Bigr\}=\Delta t \int_{-\infty}^\infty |y| p(u,y,t)dy +O(\Delta t^2).
\label{cross_prob}
\end{equation}

The average number of crossings in $(t,t+\Delta t)$ is thus 
$$
\bigl\langle N_u(t,t +\Delta t)\bigr\rangle= 1 \times {\rm Prob} \bigl\{
N_u(t,t +\Delta t) = 1\bigr\} + 0 \times {\rm Prob} \bigl\{ N_u(t,t +\Delta t) = 0\bigr\},
$$
and by virtue of Eq. \eqref{cross_prob} we write
\begin{equation}
\bigl\langle N_u(t,t + \Delta t)\bigr\rangle=\Delta t \int_{-\infty}^\infty |y| p(u,y,t)dy+ O(\Delta t^2).
\label{dN_u}
\end{equation}

We define the {\it intensity (or frequency) of crossings}, $\mu_u(t)$,  as the expected number of crossings per unit time, that is 
\begin{equation}
\mu_u(t)\equiv \lim_{\Delta t\to 0}\frac{\bigl\langle N_u(t, t+\Delta t)\bigr\rangle}{\Delta t},
\label{cross_freq}
\end{equation}
and from Eq. \eqref{dN_u} we obtain the generalized Rice formula:
\begin{equation}
\mu_u(t)=\int_{-\infty}^\infty |y| p(u,y,t)dy
\label{rice}
\end{equation}
valid for general non-stationary random processes.\footnote{As we will see below (see Eq. \eqref{rice_original}), the term ``Rice formula'' is usually applied to the case when $X(t)$ and $Y(t)$ are independent and stationary Gaussian processes with zero mean. In any case the expression \eqref{rice} is also termed as Rice formula.}
We also see from Eqs. \eqref{dN_u}--\eqref{rice} that the average $\bigl\langle N_u(t_0,t)\bigr\rangle$ of the total number of crossings during a finite time interval $(t_0,t)$ is 
\begin{equation}
\bigl\langle N_u(t_0,t)\bigr\rangle=\int_{t_0}^t \mu_u(t') dt'=\int_{t_0}^t dt' \int_{-\infty}^\infty |y| p(u,y,t')dy.
\label{N_u}
\end{equation}

Considering that the average of the total number crossings is the sum of the average number of upcrossings plus downcrossings , i.e., $\bigl\langle N_u(t_0,t)\bigr\rangle=
\bigl\langle N_u^{(+)}(t_0,t)\bigr\rangle+\bigl\langle N_u^{(-)}(t_0,t)\bigr\rangle$ (tangencies are supposed to be a set of zero measure \cite{lindgren_book}), the expressions above can be easily modified  to define the intensity of upcrossings $\mu_u^{(+)}(t)$ or downcrossings $\mu_u^{(-)}(t)$ as
\begin{equation}
\mu_u^{(+)}(t)=\int_{0}^\infty y p(u,y,t)dy,
\label{rice+}
\end{equation}
and
\begin{equation}
\mu_u^{(-)}(t)=\int_{-\infty}^0 |y| p(u,y,t)dy=\int_{0}^\infty y p(u,-y,t)dy.
\label{rice-}
\end{equation}
Obviously,
\begin{equation}
\mu_u(t)=\mu_u^{(+)}(t)+\mu_u^{(-)}(t).
\label{total_mu}
\end{equation}

An alternative way to deduce the above results is via the Kac counting
formula \cite{kac}. In order to derive this formula, following Ref.\cite{adler},
let $s_1, s_2, \dots $ be the crossing times of the $N_u(t,t_0)$ crossings of level $u$ in the interval $[t_0,t]$. Consider a sufficiently small interval $I_i$ around the crossing time $s_i$, so that no other crossings occur in this interval. Then, applying the change of variables $z=X(t)$ to the identity 
\bd
1 = \int_{-\infty}^\infty \delta(z - u) dz \,,
\ed
we obtain
\bd
1= \int_{I_i} \delta(X(t)-u) |\dot{X}(t)| dt \,,
\ed
and summing over all the crossings we obtain the celebrated Kac counting formula \cite{kac}
(in physicists' notation):
\bd
N_u(t,t_0) = \int_{t_0}^t \delta(X(t^\prime)-u) |\dot{X}(t^\prime)| dt^\prime\,.
\ed
The expectation value of $N_u(t,t_0)$ is thus
\bd
\langle N_u(t,t_0) \rangle = \int_{t_0}^t dt^\prime \int_{-\infty}^\infty dx \int_{-\infty}^\infty dy p(x,y,t^\prime) \delta(x-u) |y|
=  \int_{t_0}^t dt^\prime \int_{-\infty}^\infty p(u,y,t^\prime) |y|  dy \,.
\ed
For a rigorous derivation, we refer to \cite{adler} (p.265).
Generalizations of the Kac formula (sometimes called Kac-Rice formula) were later obtained 
for scalar-valued random fields
($X(t)\in \mathbb{R}$ and $t \in \mathbb{R}^d$ with $d>1$) \cite{longuet}, 
as well as vector-valued
random fields ($X(t)\in \mathbb{R}^{d^{\prime}}$ and $t \in \mathbb{R}^d$, generally with $d^\prime < d$). 
Moreover, extensions to the counting of critical points were also obtained.
For rigorous recent reviews of these developments, we refer to the books \cite{azais, berzin}. 
In this work, we will only be concerned with one-dimendional random processes ($d=d^{\prime}=1$).
The extension of our results to higher dimensions appears rather difficult due to the increasing complexity of the geometry.

\subsection{Stationary processes. Return time and maximum distribution}

We now suppose that $X(t)$ is a stationary random process, which means that it is time homogeneous and that there exists a time-independent stationary distribution defined as \cite{maso_llibre}
$$
p_{\rm st}(x,y)=\lim_{t\to\infty} p(x,y,t).
$$
This leads us to define the stationary intensity of crossings by
$$
\mu_u\equiv\lim_{t\to\infty} \mu_u(t).
$$
Taking the limit $t\to\infty$ in Eq. \eqref{rice}, Rice formula now reads
\begin{equation}
\mu_u=\int_{-\infty}^\infty |y| p_{\rm st}(u,y)dy,
\label{rice_stat}
\end{equation}
and the average for the total number of crossings over a finite time interval $\Delta t=t-t_0$ is given by (cf. Eq. \eqref{N_u})
\begin{equation}
\bigl\langle N_u(t_0,t_0+\Delta t)\bigr\rangle=\mu_u \Delta t=\Delta t \int_{-\infty}^\infty |y| p_{\rm st}(u,y)dy.
\label{N_stat}
\end{equation}

These expressions can be trivially extended to upcrossings and downcrossings. We thus have
$$
\mu_u^{(+)}=\int_{0}^\infty y p_{\rm st}(u,y)dy, \qquad \quad \mu_u^{(-)}=\int_{0}^\infty y p_{\rm st}(u,-y)dy. 
$$
Related to the stationary intensity of upcrossings is the {\it return period} $T_u$ to a level $u$, defined as
\begin{equation}
T_u=\frac{1}{\mu_u^{(+)}},
\label{T}
\end{equation}
which provides the mean time interval between successive upcrossings of the level $u$.

Let us next briefly explain the connection between crossing counting and the distribution of the maximum value taken by a random process $X(\tau)$ on a given time interval $\tau\in (t_0,t)$. We introduce such a connection through an engineering example. 
The return period is a key quantity in engineering for designing the maximal load that a mechanical structure can withstand before suffering structural damage, as well as for knowing its operative life \cite{rychlik}. 
Designers want to know the probability that the structure will suffer a load surpassing the design load $u$ during a certain service time $t$. Thus, if $X(\tau)$ represents the load at time $\tau$ and  
$$
M(t_0,t)= {\rm max}\{X(\tau),\  t_0\leq \tau \leq t\}
$$
is the maximum load within the service time, $(t_0,t)$, we want to know ${\rm Prob}\{M(t_0,t)>u\}$. There is a very close relation between this probability and the probability 
${\rm Prob}\{N^{(+)}_u(t)>0\}$ that there has been at least one upcrossing to level $u$ during the interval $(t_0,t)$. Indeed, assuming that the process starts below the critical value, $X_0(t_0)=x_0<u$, we have
\begin{equation}
{\rm Prob}\{M(t_0,t)>u\}={\rm Prob}\{N_u^{(+)}(t)>0\},
\label{p_max}
\end{equation}
which connects two aspects of the level-crossing problem as are extreme values and level-crossing counting.

Such a connection can be further enhanced in the following way. Let us first note that 
$$
{\rm Prob}\{M(t_0,t)>u\}=1-{\rm Prob}\{M(t_0,t)\leq u\},
$$
but ${\rm Prob}\{M(t_0,t)\leq u\}$ is the distribution function of the maximum, that is
$$
F(u,t|x_0,t_0)={\rm Prob}\{M(t_0,t)\leq u | X(t_0)=x_0\}.
$$
However, $F(u,t|x_0,t_0)$ is related to the survival (or non-hitting) probability $S$ at time $t$ of the process $X(\tau)$,
$$
S(u,t | x_0,t_0)={\rm Prob} \{X(\tau)\neq u;\ \forall \tau \in (t_0, t) |\ X(t_0)=x_0\},
$$
which is instrumental in first-passage problems. Indeed, as we have shown (see, for instance,  \cite{maso_llibre,maso_2014a,maso_2014b}) 
$$
F(u,t|x_0,t_0)=S(u,t | x_0,t_0) \Theta(u-x_0),
$$
($\Theta(\cdot)$ is the Heaviside step function) and since we have assumed that $x_0<u$ we simply write
$$
F(u,t|x_0,t_0)=S(u,t | x_0,t_0).
$$
In other words
$$
{\rm Prob}\{M(t_0,t)\leq u\}=S(u,t | x_0,t_0),
$$
and from Eq. \eqref{p_max} we write
\begin{equation}
{\rm Prob}\{N_u^{(+)}(t)>0\}=1-S(u,t | x_0,t_0) 
\label{survival}
\end{equation}
which clearly shows the relationship between first-passage (via survival probability) and level-crossing counting. For diffusion processes the survival probability can be obtained by solving the  Fokker-Planck equation with initial and absorbing boundary conditions \cite{maso_llibre} and this can provide a way of obtaining the exact expression of the probability ${\rm Prob}\{N_u^{(+)}(t)>0\}$ which is, in general, rather difficult to get \cite{rychlik}.

Let us finally obtain a practical bound for  ${\rm Prob}\{M(t_0,t)>u\}$ which may be relevant in applications. From the Markov inequality we have 
$$
{\rm Prob}\{N_u^{(+)}(t)>0\}\leq \langle N_u^{(+)}(t)\rangle \qquad \Rightarrow \qquad {\rm Prob}\{M(t_0,t)>u\}\leq \langle N_u^{(+)}(t)\rangle,
$$
and for stationary processes we write (cf. Eqs. \eqref{N_stat})  
$$
\langle N_u^{(+)}(t)\rangle=\mu_u^{(+)} \Delta t  \qquad \Rightarrow \qquad {\rm Prob}\{M(t_0,t)>u\}\leq \mu_u^{(+)} \Delta t
$$
$(\Delta t=t-t_0)$ and using Eq. \eqref{T} we have
$$
{\rm Prob}\{M(t_0,t)>u\}\leq \frac{\Delta t}{T_u},
$$
which is a useful bound for the probability that the maximum load exceeds the critical level during the time interval $\Delta t$.

\subsection{The original Rice formula}

As mentioned in the introduction, Rice's formula for level crossings was first obtained for stationary Gaussian processes, assuming that the process $X(t)$ and its derivative $Y(t)=\dot X(t)$ are uncorrelated and, hence, independent.\footnote{Recall that stationarity means that the joint PDF, $p(x,y,t)=p(x,y)$, does not depend of time, which in particular implies that the averages $\langle X(t)\rangle=m_x$ and $\langle Y(t)\rangle=m_y$ do not depend on time either and that $\langle X(t+\tau)\dot X(t)\rangle=\langle X(\tau)\dot X(0)\rangle$ for all $\tau$ and $\langle X(t)\dot X(t)\rangle=\langle X(0)\dot X(0)\rangle$. On the other hand, uncorrelated implies that 
$\langle X(\tau)\dot X(0)\rangle=\langle X(\tau)\rangle\langle\dot X(0)\rangle$ and, in particular  $\sigma^2_{xy}=\langle X(0)\dot X(0)\rangle-\langle X(0)\rangle\langle \dot X(0)\rangle=0$. Since Gaussian processes are determined by the first two moments, then uncorrelated (i.e., $\sigma_{xy}^2=0$) it also means being independent.} In such a case the joint PDF will be given by $p(x,y,t)=p(x)p(y)$, that is, 
\begin{equation}
p(x,y)=\frac{1}{2\pi\sigma_x\sigma_y} \exp\left\{-\frac{(x-m_x)^2}{2\sigma_x^2}-\frac{(y-m_y)^2}{2\sigma_y^2}\right\},
\label{pdf_gauss0}
\end{equation}
where $m_x$, $m_y$ are the stationary averages and $\sigma_x^2$, $\sigma_y^2$ the stationary variances of $X(t)$ and $\dot X(t)$ respectively.

In the original formulation it is also assumed that velocity has zero mean, i.e., $m_y=0$, then substituting Eq. \eqref{pdf_gauss0} into Eq. \eqref{rice} we readily obtain the classical Rice formula for the intensity of crossing the level $u$:
\begin{equation}
\mu_u=\frac{\sigma_y}{\pi\sigma_x} e^{-(u-m_x)^2/2\sigma_x^2}.
\label{rice_clas}
\end{equation}
When we set $u=m_x$ --corresponding to the crossing of the mean value-- we get
\begin{equation}
\mu_m =\frac{\sigma_y}{\pi\sigma_x},
\label{rice_original}
\end{equation}
which agrees with the zero-crossing intensity originally devised by Rice \cite{rice}.

\section{Level-crossing counting for general Gaussian processes}
\label{sec_3}

We have seen that Rice formula is usually written for stationary Gaussian processes $X(t)$ and when $\dot X(t)$ has zero mean and is independent of $X(t)$
(cf. Eq. \eqref{rice_clas}). Before specifically addressing inertial processes we will present Rice formula for any general Gaussian process with no restrictions. Let us thus suppose that $X(t)$ is a Gaussian process, then its derivative,  $\dot X(t)=Y(t)$, is also Gaussian since the derivative is a linear operation on $X(t)$ and keeps the Gaussian character. In its more general form the joint 
PDF of the bidimensional process $(X(t),Y(t))$ is explicitly given by  the Gaussian function \cite{maso_llibre}
\begin{equation}
p(x,y,t)=\frac{1}{2\pi\Delta(t)}\exp\biggl\{-\frac{1}{2\Delta^2(t)}\Bigl[\sigma_y^2(t)(x-m_x(t))^2-2\sigma_{xy}(t)(x-m_x(t))(y-m_y(t))+\sigma_x^2(t)(y-m_y(t))^2\Bigr]\biggr\}, 
\label{pdf}
\end{equation}
where
\begin{equation}
m_x(t)=\langle X(t)\rangle, \qquad m_y(t)=\langle Y(t)\rangle,
\label{averages}
\end{equation}
\begin{equation}
\sigma_x^2(t)=\left\langle\left[X(t)-m_x(t)\right]^2\right\rangle,\quad 
\sigma_{xy}(t)=\Bigl\langle [X(t)-m_x(t)][Y(t)-m_y(t)]\Bigr\rangle,\quad 
\sigma_y^2(t)=\left\langle\left[Y(t)-m_y(t)\right]^2\right\rangle,
\label{variances}
\end{equation}
are mean values and variances, and the discriminant $\Delta(t)$ (not to be confused with the time increment $\Delta t$ used earlier) is
\begin{equation}
\Delta(t)=\sqrt{\sigma_x^2(t)\sigma_y^2(t)-\sigma_{xy}^2(t)}.
\label{Delta_t}
\end{equation}

The total crossing intensity $\mu_u(t)$ will be given by Rice formula after substituting Eq. \eqref{pdf} into Eq. \eqref{rice}. 
We will first evaluate the intensities of upcrossings and downcrossings, $\mu_u^{(+)}(t)$ and $\mu_u^{(-)}(t)$ respectively and then obtain the total frequency 
$\mu_u(t)$. From Eqs. \eqref{rice+} and \eqref{pdf} we write
\begin{eqnarray}
\mu_u^{(+)}(t)&=&\int_0^\infty y p(u,y,t)dy \nonumber\\
&=&\frac{1}{2\pi\Delta} e^{-\sigma_y^2(u-m_x)/2\Delta^2} \int_0^\infty y \exp\left\{-\frac{\sigma_x^2}{2\Delta^2}(y-m_y)^2+
\frac{\sigma_{xy}(u-m_x)}{\Delta^2}(y-m_y)\right\}dy,
\label{mu+_1}
\end{eqnarray}
which, after performing the Gaussian integral and simple manipulations, yields  
\begin{equation}
\mu_u^{(+)}(t)=\frac{\Delta(t)}{2\pi\sigma_x^2(t)} e^{-(u-m_x(t))^2/2\sigma^2_x(t)}\left[e^{-\eta_u^2(t)} + \sqrt\pi\eta_u(t){\rm Erfc} \bigl[-\eta_u(t)\bigr]\right],
\label{mu+_gauss}
\end{equation}
where
\begin{equation}
\eta_u(t) \equiv \frac{m_y(t)\sigma_x(t)}{\sqrt 2 \Delta(t)}+\frac{\sigma_{xy}(t)}{\sqrt 2 \Delta(t)\sigma_x(t)}[u-m_x(t)],
\label{eta}
\end{equation}
and 
$$
{\rm  Erfc}(z)=\frac{2}{\sqrt\pi}\int_z^\infty e^{-t^2} dt
$$
is the complementary error function. 

As to downcrossings, from Eqs. \eqref{rice-} and \eqref{pdf} we have
\begin{eqnarray}
\mu_u^{(-)}(t)&=&\int_0^\infty y p(x,-y,t)dy \nonumber\\
&=&\frac{1}{2\pi\Delta} e^{-\sigma_y^2(u-m_x)/2\Delta^2} \int_0^\infty y \exp\left\{-\frac{\sigma_x^2}{2\Delta^2}(y+m_y)^2-
\frac{\sigma_{xy}(u-m_x)}{\Delta^2}(y+m_y)\right\}dy,
\label{mu-_1}
\end{eqnarray}
and by comparing Eq. \eqref{mu+_1} with Eq. \eqref{mu-_1} we see that, knowing $\mu_u^{(+)}(t)$ we can recover $\mu_u^{(-)}(t)$ after  making the replacements
$$
m_y(t) \longrightarrow -m_y(t), \qquad \sigma_{xy}(t) \longrightarrow -\sigma_{xy}(t).
$$
As a result from Eq. \eqref{mu+_gauss} we get
\begin{equation}
\mu_u^{(-)}(t)=\frac{\Delta(t)}{2\pi\sigma_x^2(t)} e^{-(u-m_x(t))^2/2\sigma^2_x(t)}\left[e^{-\eta_u^2(t)} - \sqrt\pi\eta_u(t){\rm  Erfc} \bigl[\eta_u(t)\bigr]\right],
\label{mu-_gauss}
\end{equation}
with $\eta_u(t)$ given in Eq.\eqref{eta}. 

The total number of crossings is given by the sum (cf. Eq. \eqref{total_mu}) 
$$
\mu_u(t)=\mu_u^{(+)}(t)+\mu_u^{(-)}(t).
$$
Adding Eqs. \eqref{mu+_gauss} and  \eqref{mu-_gauss} and taking into account that
$$
{\rm  Erfc} (-z) - {\rm  Erfc} (z) = 2 {\rm  Erf} (z),
$$
where
$$
{\rm  Erf}(z)=\frac{2}{\sqrt\pi}\int_0^z e^{-x^2}dx,
$$
is the error function, we obtain
\begin{equation}
\mu_u(t)=\frac{\Delta(t)}{\pi\sigma_x^2(t)} e^{-(u-m_x(t))^2/2\sigma^2_x(t)}\left[e^{-\eta_u^2(t)} + \sqrt\pi\eta_u(t){\rm  Erf} \bigl[\eta_u(t)\bigr]\right]\, .
\label{mu_gauss}
\end{equation}
Equations \eqref{mu+_gauss}, \eqref{mu-_gauss} and \eqref{mu_gauss} constitute the most general forms of Rice formula for any Gaussian process. 

Let us finish this section by presenting two particular but important cases. 

(i) In the first case we suppose that $X(t)$ and $Y(t)$ are independent, in which case
$$
\sigma_{xy}(t)=0 \qquad \Rightarrow \qquad \Delta(t)=\sigma_x(t)\sigma_y(t)
$$
and Eq. \eqref{mu_gauss} reads
\begin{equation}
\mu_u(t)=\frac{\sigma_y(t)}{\pi\sigma_x(t)} e^{-(u-m_x(t))^2/2\sigma^2_x(t)}\left\{e^{-m_y^2(t)/2\sigma_y^2(t)}+
\left(\frac{\pi}{2}\right)^{1/2} \frac{m_y(t)}{\sigma_y(t)} {\rm Erf} \left[\frac{m_y(t)}{\sqrt 2\sigma_y(t)}\right]\right\}.
\label{mu_indep}
\end{equation}
If, in addition, $m_y(t)=0$, we have
\begin{equation}
\mu_u(t)=\frac{\sigma_y(t)}{\pi\sigma_x(t)} e^{-(u-m_x(t))^2/2\sigma^2_x(t)},
\label{mu_indep_rice}
\end{equation}
which coincides with the  Rice original formula \eqref{rice_clas} in the stationary case when $\sigma_x$, $\sigma_y$ and $m_x$ are time-independent.

(ii) A second and more relevant case consists in counting the crossing of the mean value of the process, regardless whether $X(t)$ and $\dot X(t)$ are correlated or not. In such a case (which is, in fact, equivalent to the zero-crossing problem and will be referred to as mean-crossing problem from now on) we have
$$
u=m_x(t) \qquad \Rightarrow \qquad \eta_u(t)=\frac{m_y(t)\sigma_x(t)}{\sqrt 2 \Delta(t)} 
$$
and Eq. \eqref{mu_gauss} reads
\begin{equation}
\mu_{m}(t)=\frac{\Delta(t)}{\pi\sigma_x^2(t)}\left[e^{-m_y^2(t)\sigma_x^2(t)/2\Delta^2(t)} + \left(\frac{\pi}{2}\right)^{1/2} \frac{m_y(t)\sigma_x(t)}{\Delta(t)}
{\rm Erf} \left[\frac{m_y(t)\sigma_x(t)}{\sqrt 2 \Delta(t)}\right]\right],
\label{mu_0}
\end{equation}
where we use the notation
\begin{equation}
\mu_m(t)=\mu_{m_x(t)}(t),
\label{notation}
\end{equation}
for the crossing of the mean value. Finally, if the average velocity is zero, $m_y(t)=0$, we get
\begin{equation}
\mu_{m}(t)=\frac{\Delta(t)}{\pi\sigma_x^2(t)},
\label{mu_0_a}
\end{equation}
or more explicitly (cf. Eq. \eqref{Delta_t})
\begin{equation}
\mu_{m}(t)=\frac{ \sigma_y(t)}{\pi \sigma_x(t)} \sqrt{1-\left[\sigma_{xy}(t)/\sigma_x(t)\sigma_y(t)\right]^2},
\label{mu_0_b}
\end{equation}
which can be regarded as the generalization of the original Rice formula \eqref{rice_original} for the zero-crossing problem in the case when $X(t)$ and $\dot X(t)$ are correlated (i.e., $\sigma_{xy}(t)\neq 0$).

\section{Gaussian inertial processes. First examples}
\label{sec_4}

In many physical applications one frequently runs into random processes whose time evolution is given by a second-order differential equation with the appearance of inertial terms represented by second-order derivatives. For one-dimensional processes $X(t)$ a rather general form is given by 
\begin{equation}
\ddot X=F\Bigl(t,X,\dot X, \xi(t)\Bigr),
\label{sde_gen}
\end{equation}
where $F$ is an arbitrary function and $\xi(t)$ is the input noise, a given random process which is usually modeled as Gaussian white noise. The origin of such equations typically stems from Newton's second law of motion, where $X(t)$ represents the position of a particle moving under the effects of deterministic and random forces embodied by the function $F$. A paradigmatic example is the ``noisy oscillator'', a linear (or non-linear) oscillator perturbed by random influences, either in the frequency (Kubo oscillator) or with an external random force or even with a random damping \cite{gitterman}. A simpler, yet very relevant case, is provided by the inertial Brownian motion in which $F$ is a linear function independent of $t$ and $X$. An even simpler but highly nontrivial case is given by the random acceleration process where $F=k\xi(t)$. By applying the results of the previous section we will obtain exact expressions of the crossing intensity for these linear inertial cases. In this section we address the examples of Brownian motion and random acceleration, while in the next section we deal with the noisy oscillator.\footnote{We note that any random process $X(t)$ described by a second-order differential equation such as Eq.\eqref{sde_gen} is necessarily non Markovian \cite{maso_llibre}. However if we define $Y(t)=\dot{X}(t)$, then the two dimensional random process $(X(t),Y(t))$ obeys a first-order equation (see for example the discussion after Eq.\eqref{delta_cor}), and is thus Markovian. }

Before proceeding further let us note that all examples studied are linear. That is, $F$ is a linear function and the evolution equation \eqref{sde_gen} can be written as
\begin{equation}
\ddot X + \beta \dot X + \alpha X + \gamma=k\xi(t),
\label{sde_linear}
\end{equation}
where $\alpha$, $\beta$, $\gamma$ and $k$ are usually constant parameters, although they may be functions of time as in aging processes. In any case when the input noise $\xi(t)$ is Gaussian, the linearity of Eq. \eqref{sde_linear} ensures that the output process $X(t)$ is also Gaussian. 

As is well known, in second-order equations inertial influences decay faster than damping effects, so that, as time increases $(\beta t\gg 1)$ we have $|\ddot X(t)|\ll |\beta \dot X(t)|$  \cite{bender}. In the asymptotic regime $\beta t \to \infty$, Eq. \eqref{sde_linear} reduces to a first-order equation
\begin{equation}
\beta \dot X =- \alpha X - \gamma+k\xi(t),
\label{sde_linear_1}
\end{equation}
which is the well known Ornstein-Uhlenbeck process. Let us finally remark that Rice's approach is not applicable to first-order processes driven by white noise. Indeed, in such a case the variance of $\xi(t)$ is infinite and restricting ourselves to linear processes Eq. \eqref{sde_linear_1} implies that the variance of $\dot X(t)$ is also infinite. As a result the joint density $p(x,y,t)$ does not exists and Rice's approach is meaningless.\footnote{This can be directly seen below (cf. Eq. \eqref{mu_bm_asym}) where the limit $\beta \to\infty$ results in an  infinite crossing intensity, which is absurd.}

\subsection{Brownian motion}
\label{bm}

Suppose that $X(t)$ represents the position of a Brownian particle moving inside a medium of damping constant $\beta>0$ and external random force $\xi(t)$, whose evolution equation is given by
\begin{equation}
\ddot X+\beta \dot X=k\xi(t),
\label{sde_bm}
\end{equation}
where  $\xi(t)$ is zero-mean Gaussian white noise,
\begin{equation}
\langle\xi(t)\xi(t')\rangle=\delta(t-t'),
\label{delta_cor}
\end{equation}
and $k>0$ is the noise intensity. The initial conditions are $X(0)=x_0$ and $\dot X(0)=y_0$. 

The second-order equation \eqref{sde_bm} is equivalent to the first-order system
\begin{eqnarray*}
\dot X&=&Y \\
\dot Y&=&-\beta Y + k\xi(t),
\end{eqnarray*}
whose solution reads
\begin{eqnarray}
X(t)&=& x_0+\frac{y_0}{\beta}\left(1-e^{-\beta t}\right)+\frac{k}{\beta}\int_0^t\left[1-e^{-\beta(t-t')}\right]\xi(t')dt' \label{bm_x} \\
Y(t)&=& y_0e^{-\beta t} + k\int_0^t e^{-\beta(t-t')}\xi(t')dt' \label{bm_y},
\end{eqnarray}
from which we see (using  $\langle\xi(t)\rangle=0$) that
\begin{equation}
m_x(t)=\langle X(t)\rangle=x_0+\frac{y_0}{\beta}\left(1-e^{-\beta t}\right), \qquad m_y(t)=\langle Y(t)\rangle=y_0e^{-\beta t}.
\label{average_bm}
\end{equation}

Let us observe that the Gaussian character of the input noise $\xi(t)$ and the linearity of Eqs. \eqref{bm_x} and \eqref{bm_y} (or, alternatively, the linearity of Eq. \eqref{sde_bm}) show that $X(t)$ and $Y(t)$ are Gaussian processes as well. Therefore, in order to obtain the crossing intensity $\mu_u(t)$ for the Brownian particle to cross some position $u$, we may apply the results of the previous section which, as we have seen, need the knowledge of the variances  $\sigma_x^2(t)$, $\sigma_y^2(t)$ and $\sigma_{xy}(t)$. 

In Appendix \ref{app1} we obtain 
\begin{equation}
\sigma_x^2(t)=\frac{k^2}{\beta^3}\left(\beta t-\frac 32 + 2e^{-\beta t}-\frac 12 e^{-2\beta t}\right),
\label{sigma_x_bm}
\end{equation}

\begin{equation}
\sigma_y^2(t)=\frac{k^2}{2\beta}\left(1-e^{-2\beta t}\right),
\label{sigma_y_bm}
\end{equation}

and 

\begin{equation}
\sigma_{xy}(t)=\frac{k^2}{\beta^2}\left(\frac 12 - e^{-\beta t}+\frac 12 e^{-2\beta t}\right).
\label{sigma_xy_bm}
\end{equation}

The exact expression for the crossing intensity $\mu_u(t)$ is obtained by substituting Eqs. \eqref{average_bm}-\eqref{sigma_xy_bm} into Eq. \eqref{mu_gauss}, along with the expressions for $\Delta(t)$ and $\eta_u(t)$ given by Eqs. \eqref{Delta_t} and \eqref{eta} respectively. This ends in a rather cumbersome expression which we will not write. 

As $t\to\infty$, specifically for $\beta t\gg 1$, we see that 
\begin{equation}
m_x(t) \simeq x_0+\frac{y_0}{\beta}, \qquad m_y(t) \simeq 0,
\label{asym_m}
\end{equation}
and 
\begin{equation}
\sigma_x^2(t) \simeq \frac{k^2 t}{\beta^2}, \qquad \sigma_y^2(t) \simeq \frac{k^2}{2\beta}, \qquad \sigma_{xy}(t) \simeq \frac{ k^2}{2\beta^2}, \qquad\qquad (\beta t\gg 1).
\label{asym_sigma}
\end{equation} 

The fact that $\sigma_x^2(t)$ grows linearly with time clearly shows the well-known fact that Brownian motion is not stationary. In this asymptotic case we have
$$
\Delta(t)\simeq \frac{k^2 t^{1/2}}{\sqrt 2 \beta^{3/2}}, \qquad \frac{\Delta(t)}{\sigma_x^2(t)}\simeq\left(\frac{\beta}{2t}\right)^{1/2}, 
\qquad \eta_u(t)\simeq\frac{\beta^{1/2}}{4kt} (u-m_x),
$$
and Eq. \eqref{mu_gauss} becomes
\begin{equation}
\mu_u(t)\simeq \frac 1\pi \left(\frac{\beta}{2t}\right)^{1/2} e^{-\beta^2(u-m_x)^2/2k^2 t} \biggl\{e^{-\beta(u-m_x)^2/(4 k t)^2} + 
\sqrt\pi \frac{\beta^{1/2}}{4kt}(u-m_x) {\rm Erf} \Bigl[\frac{\beta^{1/2}}{4kt}(u-m_x)\Bigr]\biggr\}, \quad (\beta t\gg 1).
\label{mu_bm_approx}
\end{equation}

Note that when $u=m_x(t)$ the mean-crossing intensity is simply given by (cf. Eq. \eqref{notation}) 
$$
\mu_m(t)\simeq \frac 1\pi \left(\frac{\beta}{2t}\right)^{1/2}, \qquad (\beta t\gg 1).
$$
This asymptotic behavior is nonetheless extensible to any crossing level. Indeed, recalling that \cite{mos}
\begin{equation}
{\rm Erf}(z)=\frac{2}{\sqrt\pi} e^{-z^2}[z+O(z^2)],
\label{erf_asym}
\end{equation}
and expanding the exponentials in \eqref{mu_bm_approx} as $\beta t\gg 1$ we easily see that 
\begin{equation}
\mu_u(t)\simeq \frac 1\pi \left(\frac{\beta}{2t}\right)^{1/2}, \qquad\qquad (\beta t\gg 1),
\label{mu_bm_asym}
\end{equation}
which is valid for any crossing level $u$. Let us note that while the crossing intensity decreases with time, the total number of crossings actually increases with time. Indeed, from Eqs. \eqref{N_u} and \eqref{mu_bm_asym} we see that the average number of crossings within the interval $(t_0,t)$ is given by ($t_0$ and $t$ large)
\begin{equation}
\langle N_u(t)\rangle = \int_{t_0}^{t} \mu_u(t') dt'\simeq\frac 1\pi (2\beta t)^{1/2}\left[1-\sqrt{t_0/t}\right].
\label{N_bm}
\end{equation}

We validate the analytical results presented above by Monte Carlo simulation of the evolution equation 
Eqs. \eqref{sde_bm}.
The simulations are carried out using the algorithm of Ref.\cite{farago}, that we describe in Appendix \ref{appendix}. 
Fig. \ref{trajectoryBM} shows examples of random trajectories with $\beta =1, k = 1$ (see also Appendix \ref{appendix} for the definition of the units of the simulation parameters) and $x_0=y_0=0$.
For each time interval $[t,t+\delta t)$ we measure $\mu_u(t)$ by averaging over a large number (typically $10^6$) of trajectories.
Fig. \ref{mu_all_levels} shows the results corresponding to the above choice of parameters, for different values of $u$, together with the analytical expression obtained by substituting Eqs. \eqref{average_bm}-\eqref{sigma_xy_bm} into Eq. \eqref{mu_gauss}.

\begin{figure}[ht]
\includegraphics[width=0.6\linewidth,angle=0]{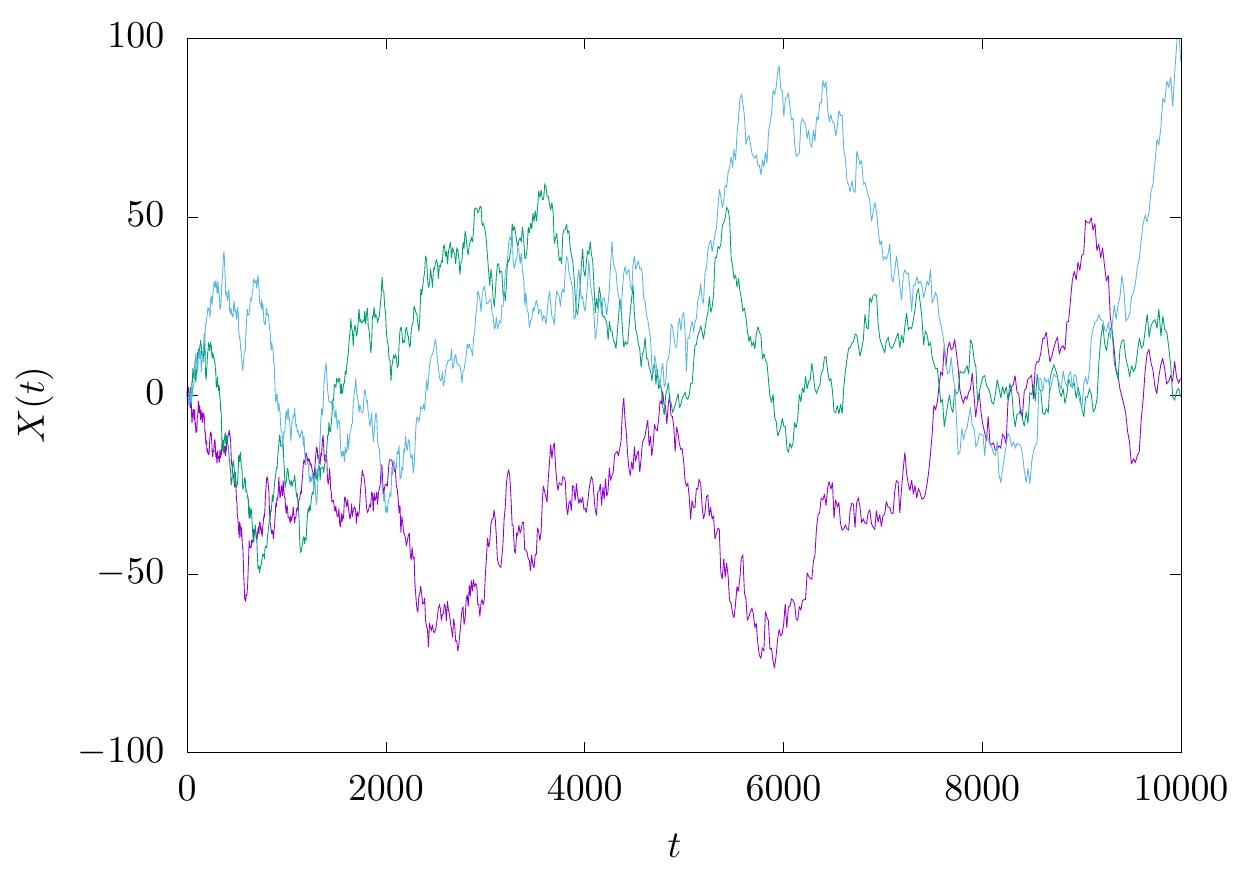}
\caption{Examples of random trajectories $X(t)$ for the Brownian motion, with $k=1, \beta=1, x_0=0, y_0=0$.
Simulations are performed with a variable time step $dt=0.01 \sqrt{t}$.}
\label{trajectoryBM}
\end{figure}

\begin{figure}[ht]
\includegraphics[width=0.6\linewidth,angle=0]{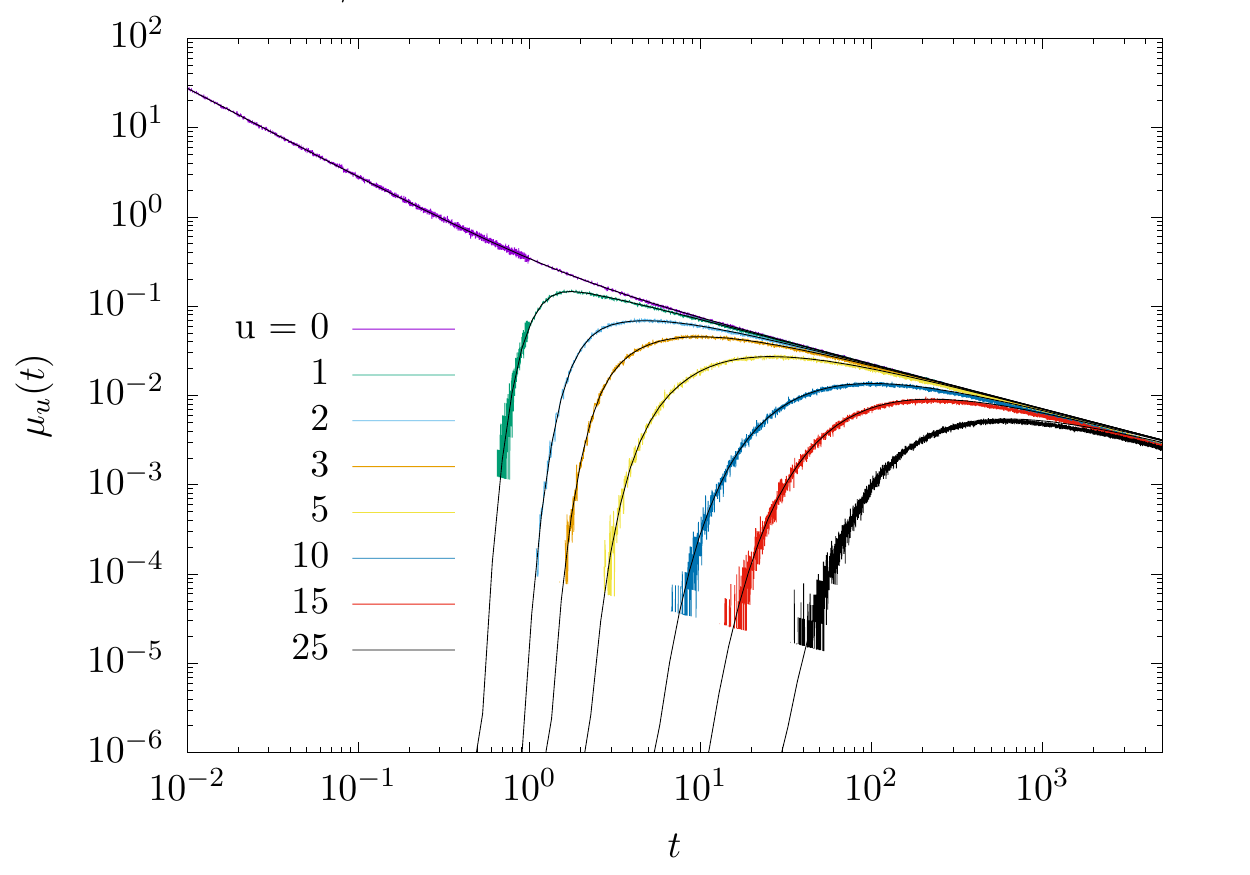}
\caption{Crossing intensity $\mu_u(t)$ for different values of $u$ obtained from simulation (noisy colored lines),
compared with the analytical prediction (smooth black lines). All simulation parameters are the same as in Fig. \ref{trajectoryBM}. The values of $u$ correspond, from top to bottom, to the lines from top to bottom.}
\label{mu_all_levels}
\end{figure}

\subsection{Random acceleration}

Let $X(t)$ be the position of an unbounded particle subject to a random acceleration represented by zero-mean Gaussian white noise 
$\xi(t)$. The dynamical equation of the process is now given by
\begin{equation}
\ddot X(t)=k\xi(t).
\label{ra}
\end{equation}
This apparently simple case represents nonetheless a nontrivial example of a non-Markovian process  and it has been the object of research in the literature related to first-exit times \cite{mas_por_96}, polymers \cite{burkhardt_07}, maxima statistics \cite{majumdar_10} and resettings \cite{singh_20} just to name a small sample. 

Denoting again $Y(t)=\dot X(t)$, and assuming $X(0)=x_0$ and $\dot X(0)=y_0$, the process, after integrating Eq. \eqref{ra}, is explicitly given by 
\begin{eqnarray}
X(t)&=& x_0+y_0 t+k\int_0^t (t-t')\xi(t')dt', \label{ra_x} \\
Y(t)&=& y_0+ k\int_0^t \xi(t')dt' \label{ra_y},
\end{eqnarray}
and 
\begin{equation}
m_x(t)=x_0+y_0 t, \qquad m_y(t)=y_0.
\label{aveerage_ra}
\end{equation}

The bidimensional process $(X(t),Y(t))$ is evidently Gaussian and proceeding as in Appendix \ref{app1} we can obtain the variances. However, since this model is a particular case of the Brownian motion after setting $\beta\to 0$, we can also obtain the variances by taking the limit $\beta\to 0$ in Eqs. \eqref{sigma_x_bm}, \eqref{sigma_y_bm} and \eqref{sigma_xy_bm}. In either way, we get
\begin{equation}
\sigma_x^2(t)=\frac 13 k^2t^3, \qquad \sigma_y^2(t)=k^2 t, \qquad \sigma_{xy}(t)=\frac 12 k^2 t^2,
\label{variances_ra}
\end{equation}
and (cf. Eqs. \eqref{Delta_t})
$$
\Delta(t)=\frac{1}{2\sqrt 3}k^2t^2.
$$

In this case the exact expression for the crossing intensity, Eq. \eqref{mu_gauss}, reads
\begin{equation}
\mu_u(t)=\frac{\sqrt 3}{2\pi t} e^{-3(u-m_x(t))^2/2k^2t^3}\left[e^{-\eta_u^2(t)}+ \sqrt\pi \eta_u(t) {\rm Erf} (\eta_u(t))\right], 
\label{mu_ra}
\end{equation}
where (cf.  Eq. \eqref{eta})
\begin{equation}
\eta_u(t)=\frac{\sqrt 2}{kt^{1/2}}\left[y_0+\frac{3}{2t}(u-m_x(t))\right].
\label{eta_ra}
\end{equation}
The mean-crossing intensity --i.e., the crossing of the mean value $u=m_x(t)=x_0+y _0t$--  is simpler and reads  
\begin{equation}
\mu_m(t)=\frac{\sqrt 3}{2\pi t} \left[e^{- 2 y_o^2/(k^2 t)}+ \frac{\sqrt{2\pi} y_0}{kt^{1/2}}{\rm Erf} \left(\frac{\sqrt{2} y_0}{kt^{1/2}}\right)\right].
\label{mu_0_ra}
\end{equation}
When $y_0=0$ we simply have
\begin{equation}
\mu_m(t)=\frac{\sqrt 3}{2\pi t}.
\label{mu_0_ra_0}
\end{equation}

Let us see next that the exact expression \eqref{mu_0_ra_0} for the mean-crossing with zero initial velocity is precisely the asymptotic expression as $t\to\infty$ of the crossing intensity for any level $u$ and any $y_0$. Indeed, from Eq. \eqref{eta_ra} we have
$$
\eta_u(t)=\frac{\sqrt 2}{kt^{1/2}}\left[y_0+O\left(\frac{1}{t}\right)\right] \qquad \Rightarrow \qquad e^{-\eta_u^2(t)}=1+O\left(\frac{1}{t}\right).
$$
Collecting results into Eq. \eqref{mu_ra}, bearing in mind that
$$
e^{-3(u-m_x(t))^2/2k^2t^3}=1+O\left(\frac{1}{t^3}\right),
$$
and recalling Eq. \eqref{erf_asym}, we finally get
\begin{equation}
\mu_u(t) \simeq\frac{\sqrt 3}{2\pi t}, \qquad (t\to\infty),
\label{mu_ra_asym}
\end{equation}
valid for any level $u$ and any initial velocity. As in the Brownian motion the crossing intensity also decreases with time, although with a different law (cf. Eq. \eqref{mu_bm_asym}), while the average number of  crossings in a time interval $(t_0,t)$ increases logarithmically ($t_0$ and $t$ large),
\begin{equation}
\langle N_u(t)\rangle = \int_{t_0}^{t} \mu_u(t') dt'\simeq\frac{\sqrt 3}{2\pi} \ln (t/t_0).
\label{N_ra}
\end{equation}

\FloatBarrier\subsection{Scaling and asymptotic regimes of the mean-crossing intensity}

We now analyze in more detail the different short- and long-time limits of the mean-crossing intensity, for both Brownian motion and random acceleration. We can identify two characteristic time scales in Brownian motion, namely 
\beq
\tau_1 = \left(\frac{y_0}{k}\right)^2 \quad {\mbox{and}}\quad \tau_2 = \beta^{-1},
\label{tau12}
\eeq
and depending on their relative value, we will obtain a different short-time behavior.

\subsubsection*{Random acceleration}

\begin{figure}[ht]
\includegraphics[width=0.6\linewidth,angle=0]{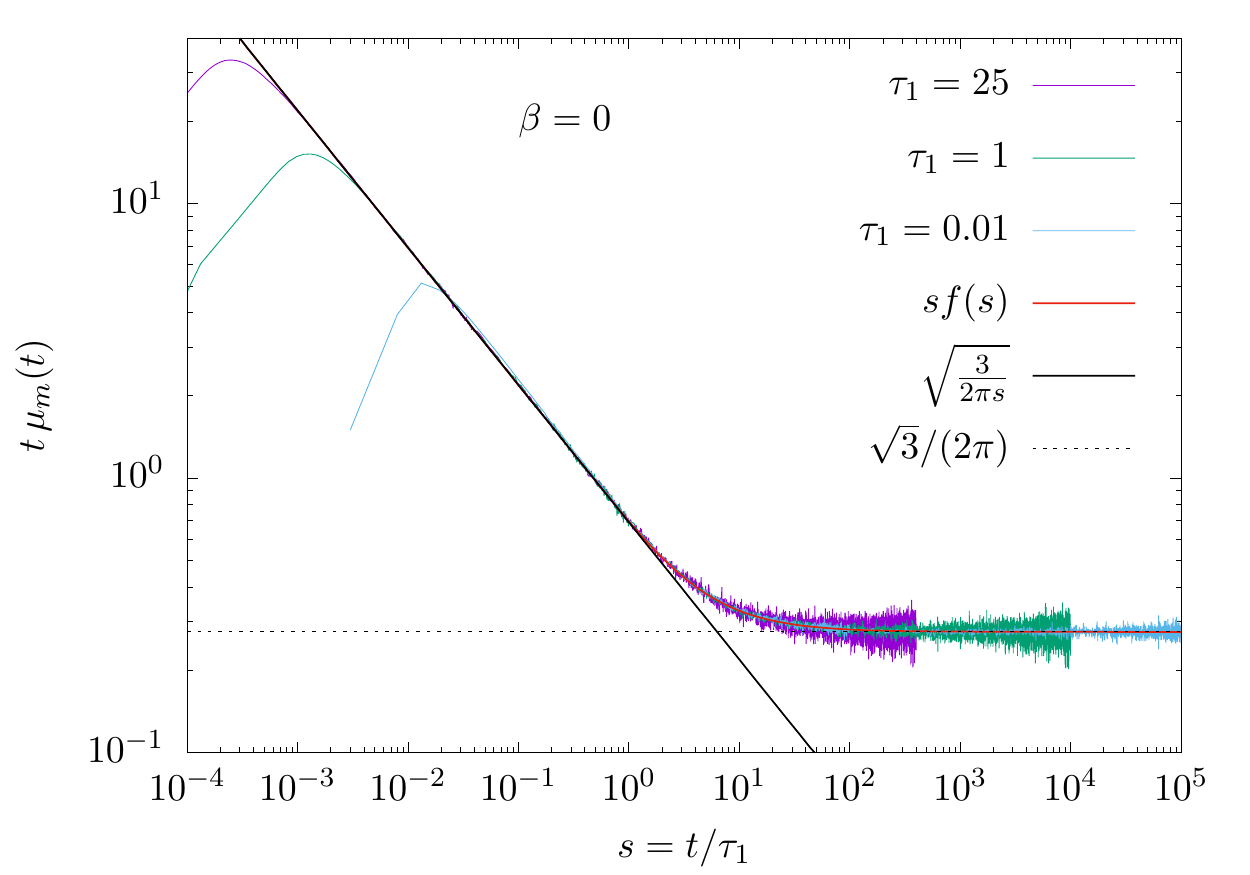}
\caption{Scaling plot of the mean-crossing intensity for random acceleration ($\beta=0$). The colored noisy lines correspond to simulation results for $k=1$ and $y_0 = 5, 1, 0.1$ (corresponding to $\tau_1 = 25, 1, 0.01$), obtained with a time step $dt = \alpha \sqrt{t}$ with $\alpha = 0.001$ for $t<1$ and $\alpha=0.01$ for $t>1$, and averaged over $10^6$ trajectories. The non-monotonic behavior at small time is an artifacts of the time discretization, which disappears 
by decreasing the time step $dt$.
The solid (red) curved line corresponds to the analytical result in Eq. \eqref{fs}, and is in perfect agreement with the simulations. 
The straight solid and dashed (black) lines correspond, respectively, to the short-time and long-time asymptotics in Eq. \eqref{limitsRA}.}
\label{scaling_y_beta0}
\end{figure}

\begin{figure}[ht]
\includegraphics[width=0.6\linewidth,angle=0]{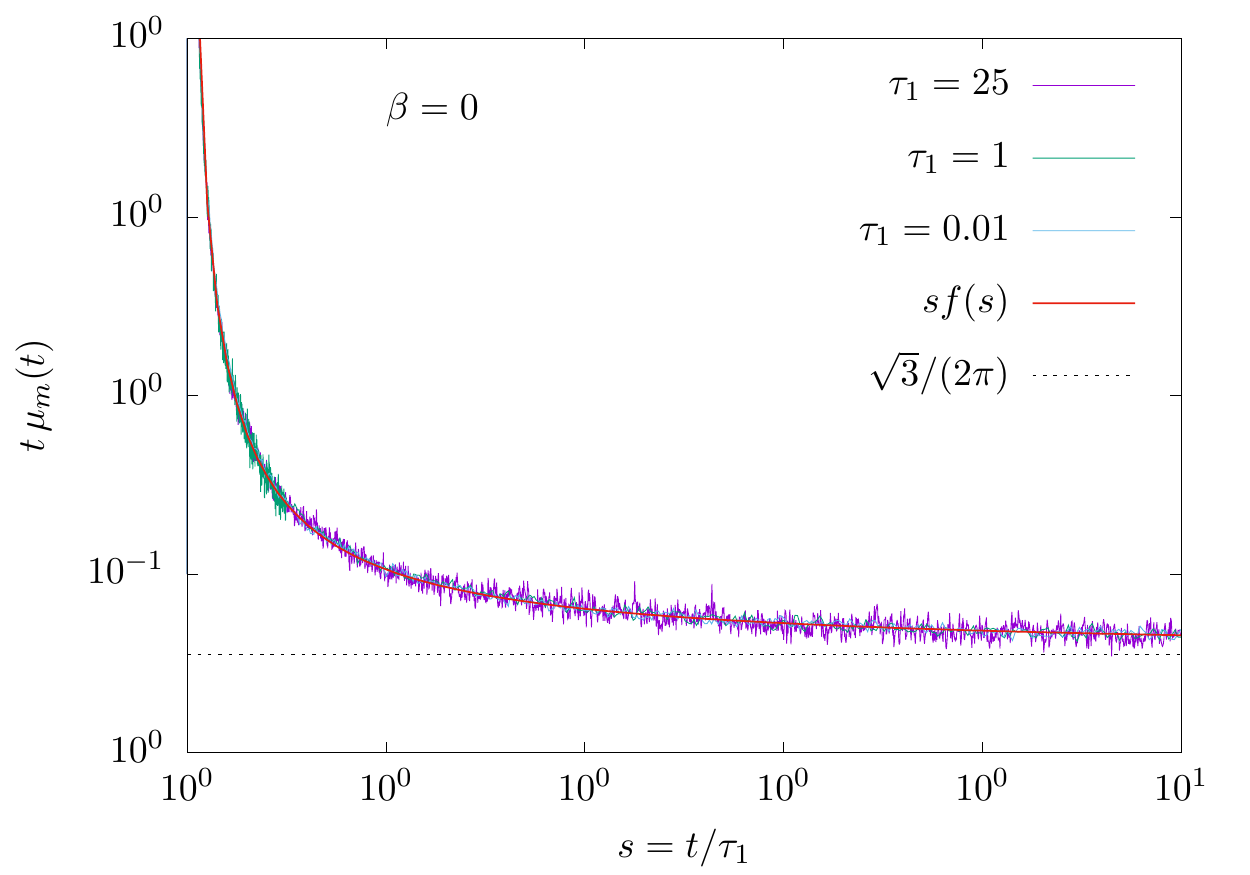}
\caption{Same as Fig. \ref{scaling_y_beta0}, but in linear scale and zooming in on the crossover region between the short- and long-time limits.}
\label{scaling_y_beta0_nolog}
\end{figure}

In this case $\beta=0$ and $\tau_2=\infty$, therefore the only relevant time scale is $\tau_1$, which is related to the initial velocity. Hence, we see from Eq. \eqref{mu_0_ra} that in this case the following scaling relation holds:
\beq
\mu_m(t) = \frac{1}{\tau_1} f(t/\tau_1)\,,
\label{eqscaling_beta0}
\eeq
where the function $f$ is given by
\beq
f(s) =  \frac{\sqrt{3}}{2 \pi s} \left[e^{-2/s} + 
\sqrt{\frac{2\pi}{s}}\, {\rm Erf}\left(\sqrt{2/s}\right)\right]\,.
\label{fs}
\eeq
The following asymptotic limits result: 
\beq  
f(s)\sim \begin{cases} \sqrt{ \frac{3}{2 \pi}}{s^{-3/2}} \quad\quad s \ll 1 \\  \\
\frac{\sqrt{3}}{2 \pi} s^{-1} \quad\quad s \gg 1 \, .
\end{cases}
\label{limitsRA}
\eeq
This scaling behavior is illustrated in Fig. \ref{scaling_y_beta0} where, in order to better appreciate the different asymptotic limits, we plot $t \mu_m(t)$, obtained from simulations at several values of $y_0$, as a function of $s = t/\tau_1$, together with the function $s f(s)$  and its asymptotic limits.
The simulation data agree perfectly with the analytical results.
An enlarged view of the crossover region at $t / \tau_1$ of order one is shown in Fig. \ref{scaling_y_beta0_nolog}.

\subsubsection*{Browmian motion}

In this case $\beta\neq 0$ and we will distinguish the cases  when the initial velocity $y_0$ is zero or different from zero. 

\begin{figure}[ht]
\includegraphics[width=0.6\linewidth,angle=0]{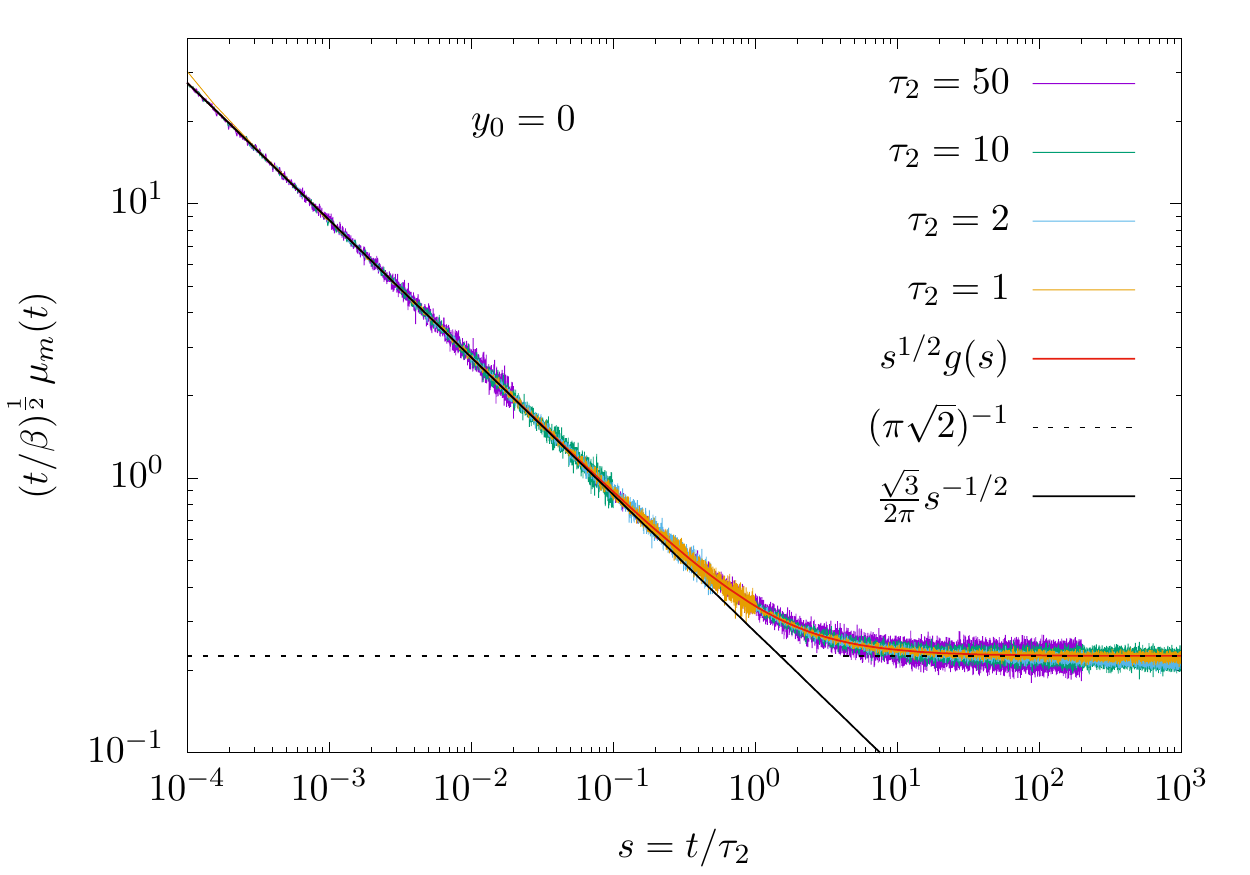}
\caption{Scaling plot of the mean-crossing intensity for $\beta\neq 0$ and $y_0=0$. 
The noisy colored lines represent simulations obtained with $\beta = 0.02, 0.1, 0.5, 1$ ($\tau_2 = 50, 10, 2, 1$). See caption of Fig. \ref{scaling_y_beta0} for details on the simulations. 
The solid (red) curved line corresponds to the analytical result in Eq. \eqref{gs}, and is in perfect agreement with the simulations. 
 The straight solid and dashed straight (black) lines correspond, respectively, to the short-time and long-time asymptotics in Eq. \eqref{limitsBM0}.}
\label{scaling_beta_y0}
\end{figure}

\begin{figure}[ht]
\includegraphics[width=0.6\linewidth,angle=0]{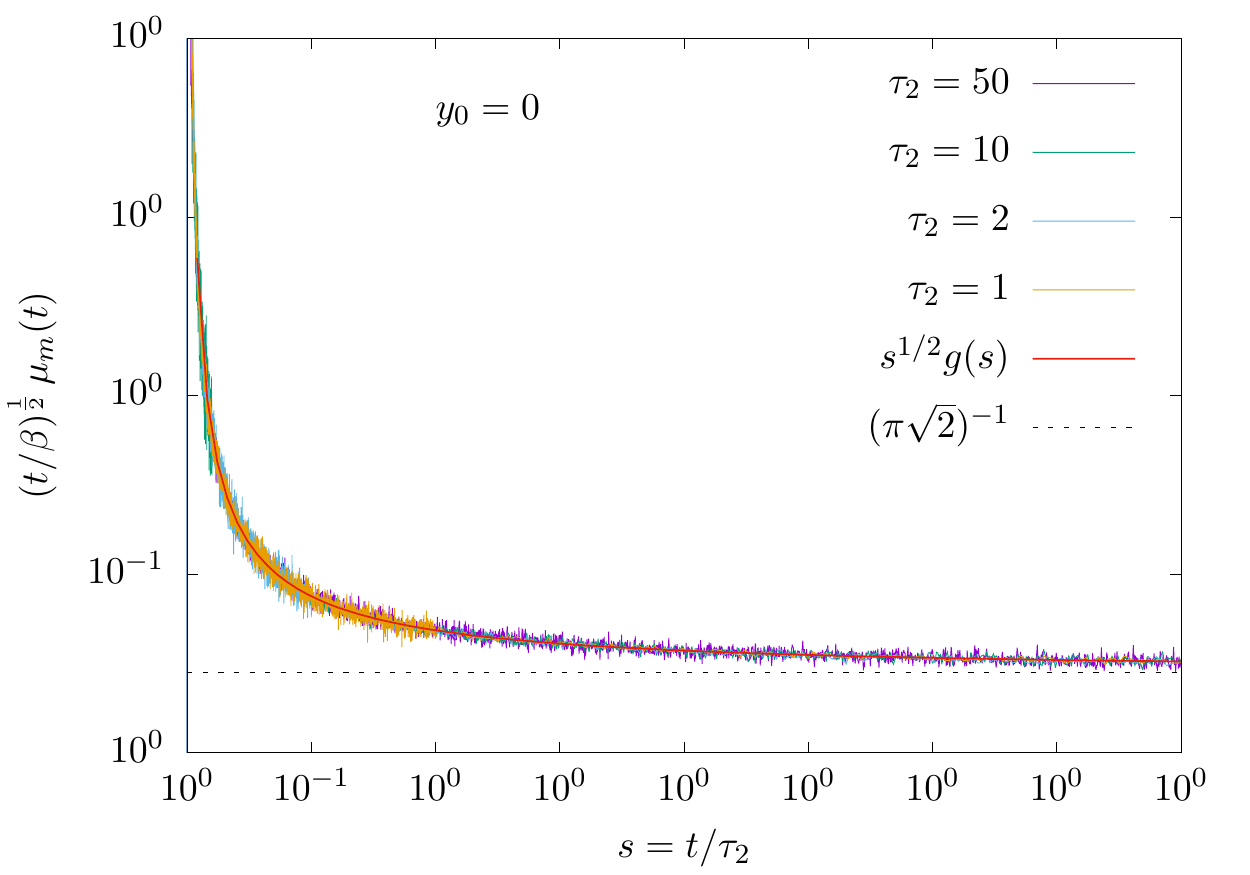}
\caption{Same as Fig. \ref{scaling_beta_y0}, but in linear scale and zooming in on the crossover region between the short- and long-time limits.}
\label{scaling_beta_y0_nolog}
\end{figure}

(i) If $y_0 = 0$ we have $\tau_1=0$ and the only relevant time scale is $\tau_2$. 
We thus see from Eqs. \eqref{mu_0_b},\eqref{sigma_x_bm},\eqref{sigma_y_bm}, and \eqref{sigma_xy_bm} that $\mu_m(t)$ satisfies a different scaling relation
\beq
\mu_m(t) = \frac{1}{\tau_2} g(t/\tau_2)
\eeq
where 
\beq
g(s)= \frac{e^s}{\pi}\frac{\left[ e^{2 s}\left( \frac{s}{2}-1\right)+2 e^s - \frac{s}{2}-1 \right]^{\frac{1}{2}}}{e^{2s}\left(s-\frac{3}{2}\right)+2 e^s -\frac{1}{2}}
\label{gs}
\eeq
and the following asymptotic limits hold:
\beq
g(s)\sim \begin{cases}
\frac{\sqrt{3}}{2 \pi} s^{-1}\quad\quad s\ll 1\\ \\
\frac{1}{\pi \sqrt{2}} {s^{-1/2}} \quad\quad s\gg 1 \,.
\end{cases}
\label{limitsBM0}
\eeq
The scaling behavior is illustrated in Figs. \ref{scaling_beta_y0} and \ref{scaling_beta_y0_nolog},  
where we plot $(t /\beta)^{1/2}  \mu_m(t)$, with $\mu_m(t)$ obtained from simulations at several values of $\beta$ and with $y_0 = 0$, as a function of $s = t/\tau_2$, together with the function $\sqrt{s} g(s)$ and its asymptotic limits.
Also in this case the simulations agree perfectly with the analytical results.

(ii) For a non-vanishing initial velocity, $y_0 \neq 0$, we have the two time scales $\tau_1$ and $\tau_2$ defined in Eq. \eqref{tau12} and from Eqs. \eqref{mu_0_b},\eqref{sigma_x_bm},\eqref{sigma_y_bm}, and \eqref{sigma_xy_bm}, we see that the crossing intensity can be written as 
\beq
\mu_m(t) = \frac{1}{\tau_2} h(t/\tau_2, \tau_2/\tau_1)
\label{scaling_general}
\eeq
where 
\beq
h(s,r) = g(s)   \,\frac{2\pi }{\sqrt{3}\, }r \, q(s) \, f[r q(s)]\,, \quad  s=t/\tau_2, \,\, r=\tau_2/\tau_1
\eeq
Here, $f$ and $g$ are the functions defined in Eqs. \eqref{fs} and \eqref{gs}, respectively, and  $q(s)$ is
the function
\beq
q(s) =  \frac{2 e^{2 s} \left(s - 2\right) + 8\, e^s -4 - 2 s}{2 e^{-s}-\frac{1}{2}e^{-2 s}+s - \frac{3}{2}} \,.
\eeq
Eq. \eqref{scaling_general} defines a family of scaling relations parametrized by the ratio $\tau_2/\tau_1$.
In the limits $\tau_1\neq 0, \tau_2\to\infty$ and $\tau_1\to 0, 0 <\tau_2 < \infty$, 
Eq. \eqref{scaling_general} reduces to, respectively, the aforementioned cases of random acceleration and Brownian motion with zero initial velocity (case (i)).

Using $q(s)\sim s$ for $s\to 0$ and  Eq. \eqref{limitsRA},  we obtain 
\bd
\mu_m(t) \sim 
\begin{cases}
\sqrt{\frac{3}{2\pi}} \frac{1}{\tau_2} 
\left(\frac{\tau_2}{t}\right)^{3/2} 
\left(\frac{\tau_1}{\tau_2} \right)^{1/2}, \quad \quad t \ll \tau_2,\\
\frac{\sqrt{3}}{2 \pi t^{1/2}}  \quad \quad t \gg (\tau_1, \tau_2)
\end{cases}
\ed
or equivalently
\beq
h(s,r) \sim \begin{cases}
\sqrt{\frac{3}{2\pi}} r^{-1/2} s^{-3/2} \quad \quad s \to 0,\\
\frac{1}{\pi \sqrt{2}} s^{-1/2}\quad \quad  s\to \infty  \,.
\end{cases}
\label{limits_general}
\eeq

\begin{figure}[ht]
\includegraphics[width=0.6\linewidth,angle=0]{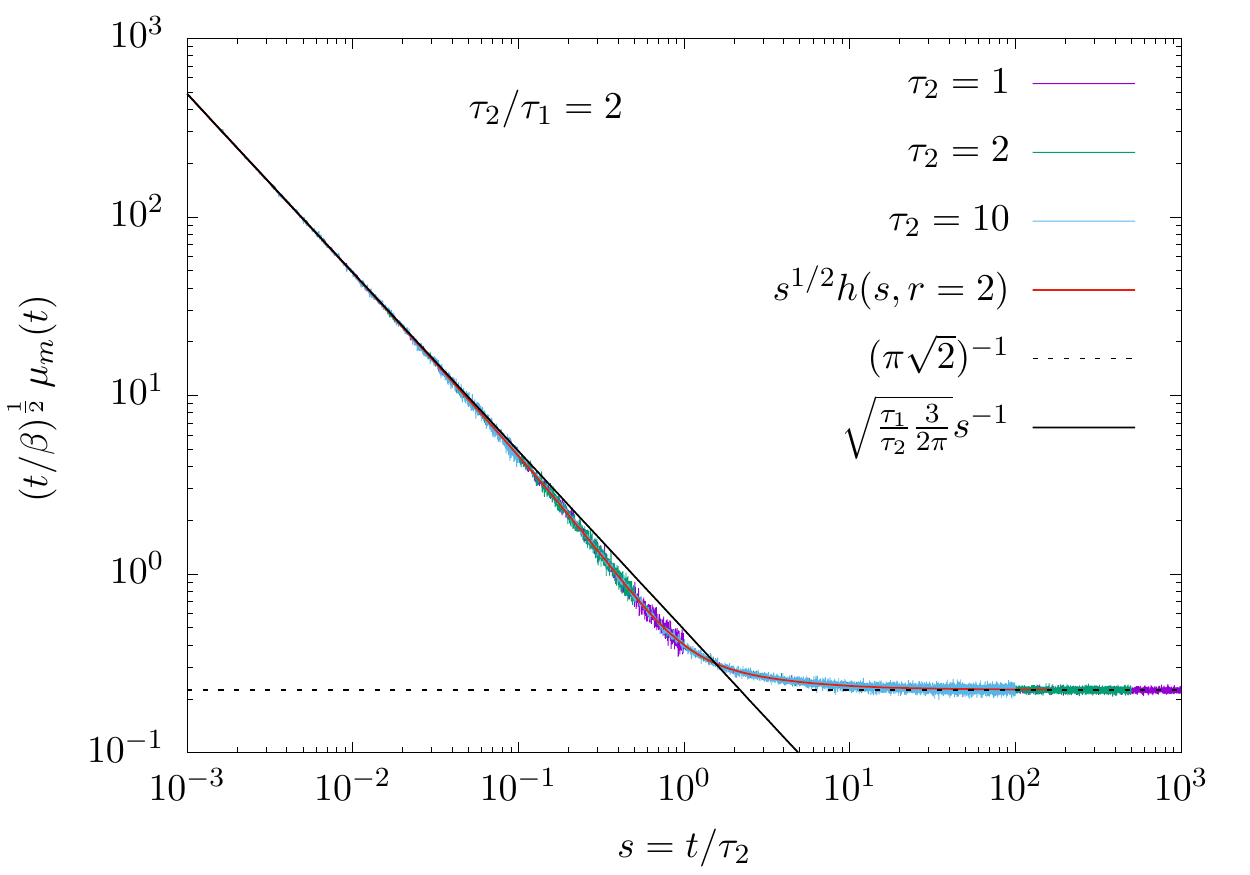}
\caption{Scaling plot of the mean-crossing intensity for $\tau_2/\tau_1=2$. The noisy colored lines correspond to the simulation results for different values of $\tau_2$. 
The curved (red) solid line corresponds to the analytical result in Eq. \eqref{scaling_general}.
 The straight solid and dashed (black) lines correspond, respectively, to the short-time and long-time asymptotic regimes in Eq. \eqref{limits_general}.}
\label{scaling_ratio2}
\end{figure}

\begin{figure}[ht]
\includegraphics[width=0.6\linewidth,angle=0]{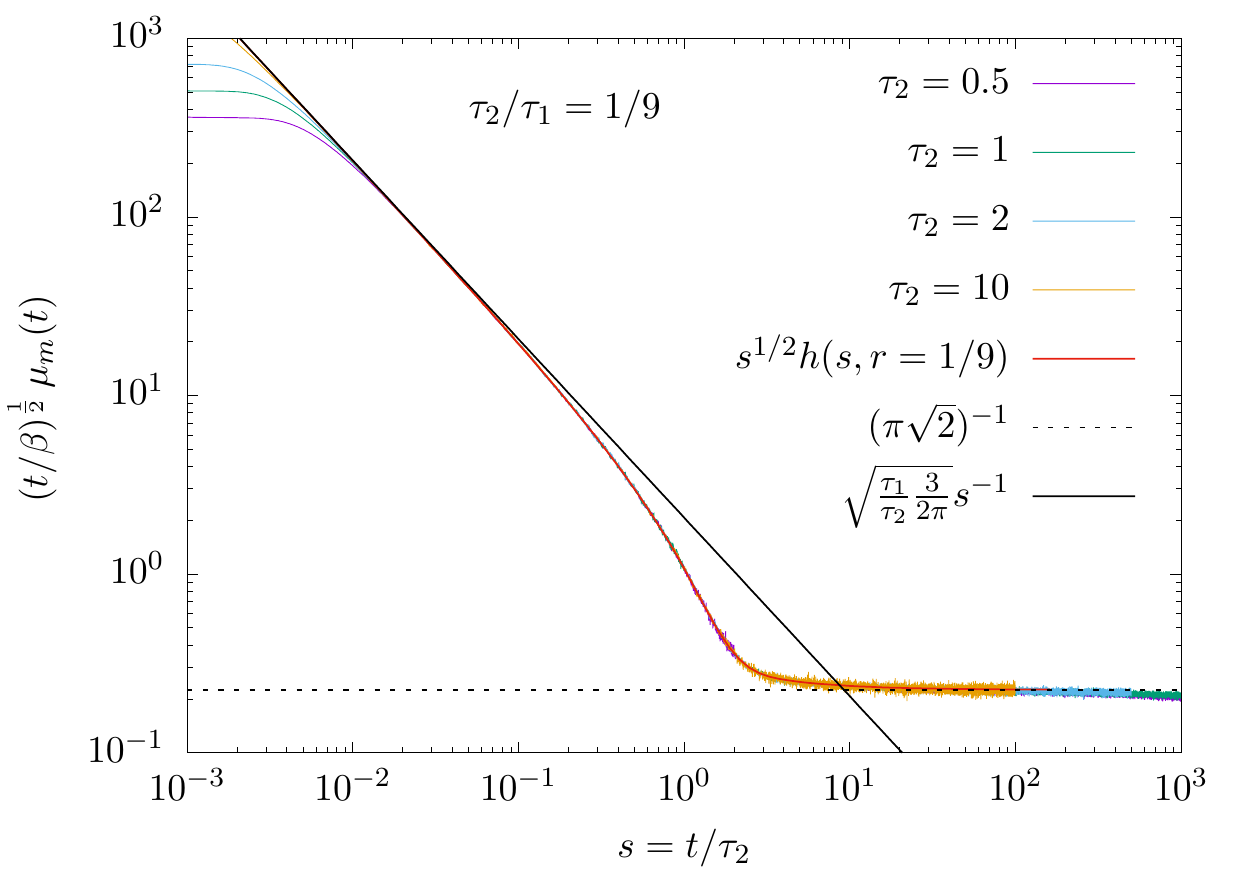}
\caption{Same as Fig. \ref{scaling_ratio2} but for $\tau_2/\tau_1=1/9$.}
\label{scaling_ratio9}
\end{figure}

\begin{figure}[ht]
\includegraphics[width=0.6\linewidth,angle=0]{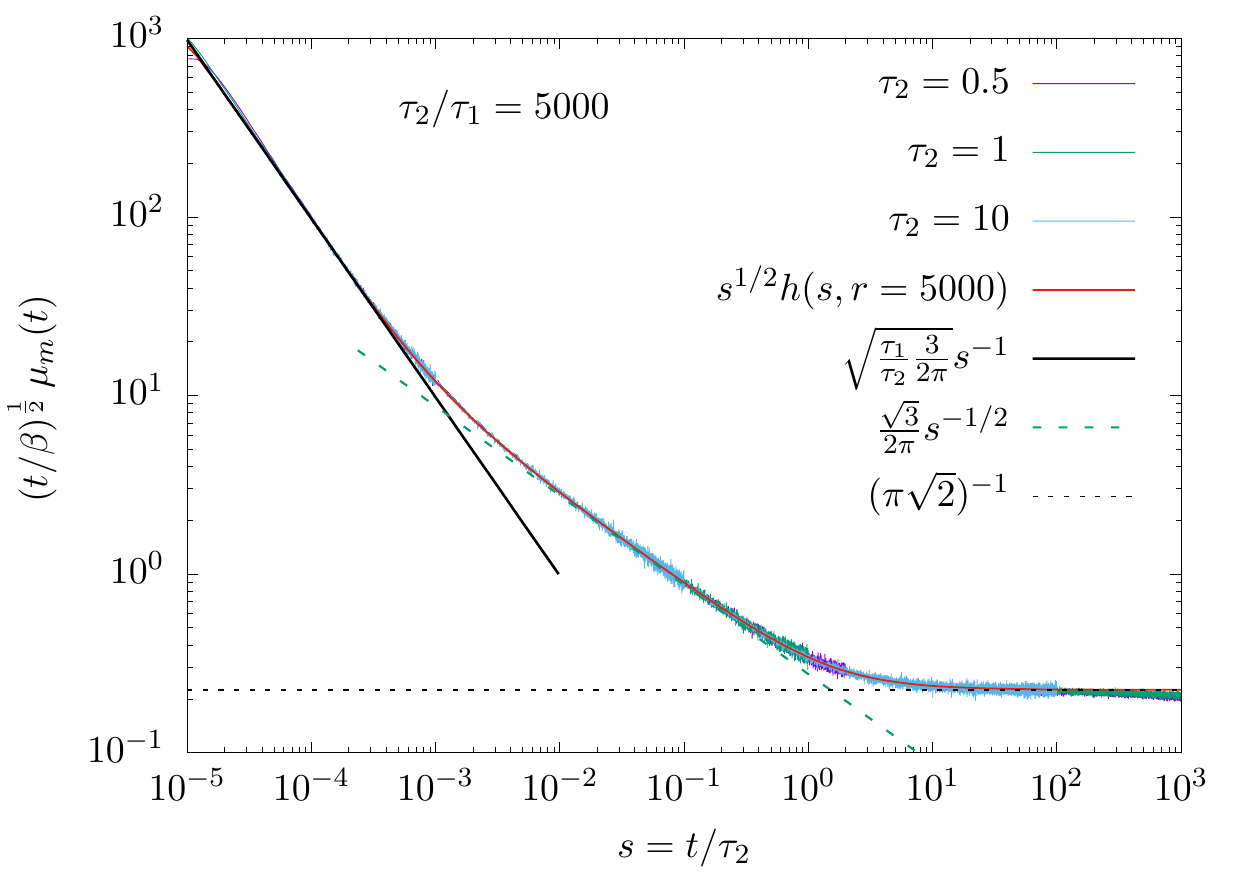}
\caption{Same as Fig. \ref{scaling_ratio2} but for $\tau_2/\tau_1=5000$.}
\label{scaling_ratio2E-4}
\end{figure}

In particular, for a given ratio $\tau_2/\tau_1$ and when $t$ is small enough we are in the ``ballistic'' regime $\mu_m(t)\sim t^{-3/2}$.  Let us note that the case 
$\tau_2/\tau_1=2$ is especially relevant since it corresponds to choosing an initial velocity equal to the asymptotic value of the mean-squared velocity. That is,
\beq
y_0^2 = \lim_{t\to \infty} \langle Y^2(t)\rangle = \frac{k^2}{2\beta},
\eeq
where we have used Eq. \eqref{sigma_y_bm}. As we will see below this is a natural choice for the initial velocity for the Brownian motion of a particle
\footnote{Let us remark that with this choice the mean-squared displacement $\langle \Delta^2 X(t) \rangle$, where $\Delta X(t) = X(t) - X(0)$,   scales as $\langle \Delta X^2(t)\rangle \simeq k^2 t^2/2 \beta$ in the ballistic regime where $t\ll \tau_2$ \cite{uhlenbeck}. This can be easily checked using Eq. \eqref{bm_x} and proceeding in the same way we obtained Eq. \eqref{sigma_x_bm}. In the diffusive regime, $t\gg \tau_2$, we have the expected diffusive behavior $\langle \Delta X^2(t) \rangle = \sigma_x^2(t) \simeq k^2 t/\beta^2$.
}.
Figures \ref{scaling_ratio2}, \ref{scaling_ratio9}, and \ref{scaling_ratio2E-4} show scaling plots for $\tau_2/\tau_1 = 2, 1/9$, and $5000$, respectively.
Notice that if the two time scales are amply separated (i.e., $1 \ll \tau_1 \ll \tau_2$) we will have three power-law regimes,  namely ballistic, random-acceleration, and diffusive:
\beq
\mu_m(t)\sim
\begin{cases}
\frac{\sqrt{3}}{2 \pi}\frac{y_0}{k t^{3/2}} \quad \quad    1 \ll t \ll \tau_1
\\ \\
\frac{\sqrt{3}}{2 \pi t} \quad \quad    \tau_1 \ll t \ll \tau_2
\\ \\
\frac{1}{\pi}\sqrt{\frac{\beta}{2 t}}\quad\quad t \gg \tau_2
\end{cases}
\eeq
as can be seen in Fig. \ref{scaling_ratio2E-4}.

It is interesting to interpret the above results in the case of the Brownian motion of a particle of mass $m$ under a viscous drag. The movement of the particle in one dimension is described by the equation
\beq
m \ddot X(t) = - \gamma \dot X(t) + \xi(t)
\eeq
where $\gamma$ is the drag coefficient (for example, $\gamma = 6 \pi \eta r$ for a spherical particle, where $\eta$ is the fluid viscosity
and $r$ is the particle radius), and $\xi$ is zero-mean Gaussian white noise satisfying the fluctuation-dissipation theorem,
\beq
\la \xi(t) \xi(t^\prime)\ra = 2 \gamma k_B T \delta(t-t^\prime),
\eeq
where $k_B$ is the Boltzmann constant and $T$ is the temperature. 
By comparison with Eq. \eqref{sde_bm}, we see that  $k = \sqrt{2 \gamma k_B T}/m$ and $\beta = \gamma/m$. Thus the duration of the ballistic regime is $\tau_2 = m / \gamma$ (a result obtained long ago by Einstein \cite{einstein}), and 
\beq
\frac{\tau_2}{\tau_1} = \frac{k_B T}{m y_0^2/2}
\eeq 
which is the ratio between twice the thermal energy and the initial kinetic energy.
In an experiment tracking the motion of an individual particle, it is natural to assume that,  when we start observing the particle, its velocity
is already thermalized, namely $y_0^2 = \lim_{t\to \infty} \langle Y^2(t)\rangle = k^2/2\beta$. Thus we have $ m y_0^2/2 = k_B T/2$ and $\tau_2/\tau_1=2$.
Furthermore, as discussed above, the mean-squared displacement behaves as $\langle \Delta X^2(t)\rangle \simeq (k_B T/m)t^2$ in the ballistic regime $t\ll m/\gamma$, and 
$\langle \Delta X^2(t)\rangle \simeq (2 k_B T/\gamma)t$ in the diffusive regime $t\gg m/\gamma$. 
The crossover between the two regimes, albeit more complex due to hydrodynamic interactions, has been observed experimentally \cite{huang}.

\FloatBarrier\section{Noisy oscillators}
\label{sec_5}

We now apply the results of Sect. \ref{sec_3} to harmonic oscillators driven by Gaussian white noise. The linearity of such systems ensures the Gaussian character of the oscillator response. We first focus on the  damped case, which is stationary, and latter address the undamped oscillator, a non-stationary process presenting some distinctive and interesting features. 

\subsection{Noisy oscillators with damping}
\label{osc_damping}

We consider a linear oscillator subject to damping and driven by an external force assumed to be zero-mean Gaussian white noise. The time evolution is given by the 
second-order linear equation 
\begin{equation} 
\ddot X+\beta\dot X+\omega_0^2 X=k\xi(t),
\label{sde_osc}
\end{equation}
where $\beta>0$ is the damping constant, $\omega_0$ is the natural frequency of the deterministic oscillator without damping, and $\xi(t)$ is Gaussian white noise with 
$\langle\xi(t)\rangle=0$ and $\langle \xi(t)\xi(t')\rangle=\delta(t-t')$. Again, due to the linearity of Eq. \eqref{sde_osc}, both $X(t)$ and $\dot X(t)$ are Gaussian processes. 

As we can see by direct substitution, the solution to Eq. \eqref{sde_osc} is 
\begin{equation}
X(t)= Ae^{-\beta(t-t_0)/2}\cos\bigl[\omega(t-t_0)+\delta\bigr] + \frac k\omega \int_{t_0}^t e^{-\beta(t-t')/2}\sin\bigl[\omega(t-t')\bigr]\xi(t') dt'.
\label{osci_x}
\end{equation}
Hence
\begin{equation}
Y(t)=\dot X(t)= -\frac{\beta}{2} X(t) - A\omega e^{-\beta(t-t_0)/2}\sin\bigl[\omega(t-t_0)+\delta\bigr] + k \int_{t_0}^t e^{-\beta(t-t')/2}\cos\bigl[\omega(t-t')\bigr]\xi(t') dt', 
\label{osci_y}
\end{equation}
where
\begin{equation}
\omega=\sqrt{\omega_0^2-\beta^2/4}.
\label{omega}
\end{equation}

In what follows we will assume that the oscillator works within the underdamped regime, i.e., $\beta<2\omega_0,$ so that $\omega$ is real. The constants $A$ and $\delta$ are related to the initial conditions, $X(t_0)=x_0$ and $\dot X(t_0)=y_0$, by
\begin{equation}
A=\sqrt{x_0^2+\frac{1}{\omega^2}(y_0+\beta x_0/2)^2}, \qquad\quad \delta=-\arctan\left[ \frac 1\omega\left(\frac{y_0}{x_0}+\frac{\beta}{2}\right)\right] \, .
\label{initial_osc}
\end{equation}
From Eqs. \eqref{osci_x}-\eqref{osci_y} we see that the average values of position and velocity are
\begin{equation}
m_x(t)=Ae^{-\beta(t-t_0)/2} \cos\bigl[\omega(t-t_0)+\delta] ,\qquad m_y(t)=-\frac{\beta}{2} m_x(t)- A\omega e^{-\beta(t-t_0)/2} \sin\bigl[\omega(t-t_0)+\delta].
\label{average_osc}
\end{equation}
Let us incidentally note that these average values correspond to the response of the deterministic oscillator.

In Appendix \ref{app2} we show that the variances are
\begin{equation}
\sigma_x^2(t)=\frac{k^2}{\omega^2(\beta^2+4\omega^2)}\left\{\frac{2\omega^2}{\beta}-
e^{-\beta(t-t_0)}\left[\beta\sin^2\omega(t-t_0)+\omega\sin 2\omega(t-t_0)+\frac{2\omega^2}{\beta}\right]\right\},
\label{sigma_x_osc}
\end{equation}
\begin{equation}
\sigma_y^2(t)=\frac{k^2}{\beta^2+4\omega^2}\left\{\frac 1\beta(\beta^2+2\omega^2)-
e^{-\beta(t-t_0)}\left[\beta\cos^2\omega(t-t_0)-\omega\sin 2\omega(t-t_0)+\frac{2\omega^2}{\beta}\right]\right\},
\label{sigma_y_osc}
\end{equation}
and
\begin{equation}
\sigma_{xy}(t)=\frac{k^2}{2\omega(\beta^2+4\omega^2)}\left\{2\omega-e^{-\beta(t-t_0)} 
\left[\beta\sin 2\omega(t-t_0)+2\omega\cos 2\omega(t-t_0)\right]\right\}.
\label{sigma_xy_osc}
\end{equation}

Knowing mean values and variances the exact expression for the crossing intensity of the oscillator to any level $u$ is attained from Eq. \eqref{mu_gauss} after using Eqs. \eqref{Delta_t} and \eqref{eta}. As in Brownian motion the resulting expression is clumsy and we will not write it explicitly. In any case the exact expression is mostly useful when the oscillator is in the transient state which may be useful in some specific applications. However,  the behavior of the oscillator at longer times, when it enters into the stationary regime, turns out to be more relevant.

Contrary to the two cases developed in the previous section which are not stationary, the noisy oscillator \eqref{sde_osc} achieves the stationary regime at long times which exclude transient effects depending on the initial conditions. This is easily seen by taking the limit $t_0\to-\infty$ in Eqs. \eqref{osci_x}-\eqref{osci_y}, that is
\footnote{Let us recall that the stationary state is achieved when $t-t_0 \to \infty$. Such a limit may be taken by two different but equivalent ways: (i)  either $t_0$ is finite (for instance $t_0=0$) and $t\to\infty$, or (ii) $t$ is finite but the process started in the infinite past, so that $t_0\to-\infty$. In writing  Eqs. \eqref{x_stat} and \eqref{y_stat} we have taken the second interpretation.}
\begin{eqnarray}
X(t)&=& \frac k\omega \int_{-\infty}^t e^{-\beta(t-t')/2}\sin\bigl[\omega(t-t')\bigr]\xi(t') dt', \label{x_stat} \\
Y(t)&=& \frac k\omega \int_{-\infty}^t e^{-\beta(t-t')/2}\bigl[-(\beta/2)\sin\omega(t-t')+\omega\cos\omega(t-t')\bigr]\xi(t') dt'. 
\label{y_stat}
\end{eqnarray}
In this regime (cf Eq. \eqref{average_osc})
\begin{equation}
m_x(t)=m_y(t)=0,
\label{stat_average}
\end{equation}
and taking the limit $t-t_0\to\infty$ in Eqs. \eqref{sigma_x_osc}-\eqref{sigma_xy_osc} we get the stationary variances 
$$
\sigma^2_x=\frac{2k^2}{\beta(\beta^2+4\omega^2)}, \qquad \sigma^2_y=\frac{k^2(\beta^2+2\omega^2)}{\beta(\beta^2+4\omega^2)}, \qquad
\sigma_{xy}=\frac{k^2}{\beta^2+4\omega^2},
$$
which, in terms of the natural frequency $\omega_0$ (cf. Eq. \eqref{omega}), can be written as
\begin{equation}
\sigma^2_x=\frac{k^2}{2\beta\omega_0^2}, \qquad \sigma^2_y=\frac{k^2(\beta^2/2+2\omega_0^2)}{4\beta\omega_0^2}, \qquad
\sigma_{xy}=\frac{k^2}{4\omega_0^2},
\label{stat_sigma}
\end{equation}
so that (cf. Eq. \eqref{Delta_t})
\begin{equation}
\Delta=\frac{k^2}{2\beta\omega_0}.
\label{stat_delta}
\end{equation}

Substituting Eqs. \eqref{stat_average}, \eqref{stat_sigma} and \eqref{stat_delta} into Eqs. \eqref{eta} and \eqref{mu_gauss}, after simple manipulations, result in the stationary crossing intensity of the noisy oscillator:
\begin{equation}
\mu_u=\frac{\omega_0}{\pi} e^{-\beta\omega_0^2 u^2/k^2}\left[e^{-\beta^3u^2/4k^2} + \sqrt\pi\frac{\beta^{3/2}u}{2k}  {\rm  Erf} \left(\frac{\beta^{3/2}u}{2k}\right)\right],
\label{stat_mu}
\end{equation}
Note that in this case crossing the mean value corresponds to setting $u=0$, which gives
\begin{equation}
\mu_m=\frac{\omega_0}{\pi},
\label{stat_mu_0}
\end{equation}
and we see that this crossing frequency (i.e., intensity) doubles the natural frequency of the deterministic oscillator.

Following the same procedure for the upcrossing and downcrossing intensities given in Eqs. \eqref{mu+_gauss} and \eqref{mu-_gauss} we easily find
\begin{equation}
\mu_u^{(\pm)}=\frac{\omega_0}{2\pi} e^{-\beta\omega_0^2 u^2/k^2}\left[e^{-\beta^3u^2/4k^2} \pm \sqrt\pi\frac{\beta^{3/2}u}{2k}  
{\rm  Erfc} \left(\mp \frac{\beta^{3/2}u}{2k}\right)\right],
\label{stat_mu+-}
\end{equation}
For the mean-crossing problem $u=m_x=0$ and both intensities are equal
$$
\mu_m^{(+)}=\mu_m^{(-)}=\mu_m/2,
$$
and the frequencies of up and down crossings equal the natural frequency of the deterministic oscillator $\omega_0/2\pi$.

\subsection{The undamped oscillator}

When no damping is present, the evolution equation of the noisy linear oscillator is simply given by
\begin{equation} 
\ddot X+\omega_0^2 X=k\xi(t).
\label{sde_osc_0}
\end{equation}
The formal solution to this equation with the initial conditions $X(0)=x_0$ and $\dot X(0)=y_0$ reads (see Eqs. \eqref{osci_x} and \eqref{osci_y})
\begin{equation}
X(t)= A\cos\big(\omega_0 t+\delta\bigr) + \frac {k}{\omega_0} \int_{0}^t \sin\omega_0(t-t')\xi(t') dt', 
\label{osci_x_0}
\end{equation}
and
\begin{equation}
Y(t)= - A\omega_0 \sin(\omega_0 t+\delta) + k \int_{0}^t \cos\omega_0(t-t')\xi(t') dt',
\label{osci_y_0}
\end{equation}
where 
\begin{equation}
A=\sqrt{x_0^2+y_0^2/\omega_0^2}, \qquad \delta=-\arctan\left(\frac{y_0}{\omega x_0}\right),
\label{A_0}
\end{equation}
and we have set $t_0=0$ without loss of generality because the process is time homogeneous, although not stationary, but obviously Gaussian. 

The average values are 
\begin{equation}
m_x(t)=A\cos(\omega_0 t+\delta) ,\qquad m_y(t)=- A\omega_0 \sin(\omega_0 t+\delta) ,
\label{average_osc_0}
\end{equation}
and variances are now given by (cf. Appendix \ref{app2})
\begin{equation}
\sigma_x^2(t)=\frac{k^2t}{2\omega_0^2}\left(1-\frac{1}{2\omega_0t} \sin 2\omega_0 t\right), \quad 
\sigma_y^2(t)=\frac{k^2t}{2}\left(1+\frac{1}{2\omega_0t} \sin 2\omega_0 t\right), \quad \sigma_{xy}(t)=\frac{k^2}{4\omega_0^2}\left(1-\cos 2\omega_0 t\right),
\label{sigma_undamped}
\end{equation}
and (cf. Eq. \eqref{Delta_t})
\begin{equation}
\Delta(t)=\frac{k^2 t}{2\omega_0} \sqrt{1-\left(\frac{\sin \omega_0 t}{\omega_0 t}\right)^2}.
\label{delta_osc_0}
\end{equation}

Substituting these expressions into Eqs. \eqref{eta} and \eqref{mu_gauss} we get the exact expression of the crossing intensity $\mu_u(t)$ for the undamped linear oscillator.  Let us, however, focus on the behavior for large times --specifically when several periods, $T_0=2\pi/\omega_0$, of the deterministic oscillator have elapsed-- that is, when 
$\omega_0 t\gg 1$. In such a case one can easily check that 
\begin{equation}
\frac{\Delta(t)}{\sigma_x^2(t)}=\omega_0\frac{\sqrt{1-\sin^2\omega_0t/\omega_0^2t^2}}{1-\sin 2\omega_0t/2\omega_0t}=\omega_0\left[1+O\left(\frac{1}{\omega_0t}\right)\right],
\label{delta/sigma}
\end{equation}
and
\begin{equation}
\eta_u(t)=\frac{1}{kt^{1/2}}\left[m_y(t)+\frac{1}{2t}(u-m_x(t)(1-\cos 2\omega_0 t))\right]\left[1+O\left(\frac{1}{\omega_0t}\right)\right].
\label{eta_osc_0}
\end{equation}

Let us incidentally note that within the same degree of approximation the function $\eta_u(t)$ is independent of the crossing level $u$ for sufficiently large values of $t$. 
Indeed, from the above expression we see that
\begin{equation}
\eta_u(t)=\frac{m_y(t)}{kt^{1/2}}\left[1+O\left(\frac{1}{\omega_0t}\right)\right],
\label{eta_assym}
\end{equation}
which is valid for all finite values of the crossing level $u$. Finally, substituting \eqref{delta/sigma} and \eqref{eta_assym} into \eqref{mu_gauss} and taking into account (cf. Eq. \eqref{sigma_undamped}) that
$$
\sigma_x^2(t)=\frac{k^2 t}{2\omega_0^2}\left[1+O\left(\frac{1}{\omega_0 t}\right)\right], 
$$
we have
$$
\mu_u(t)=\frac{\omega_0}{\pi}e^{-\omega_0^2[u-m_x(t)]^2/k^2t}\left[e^{-m_y^2(t)/k^2t}+\sqrt\pi\frac{m_y(t)}{kt^{1/2}}{\rm Erf} \left(\frac{m_y(t)}{kt^{1/2}}\right)+ 
O\left(\frac{1}{\omega_0t}\right)\right].
$$
Recalling the asymptotic expression \eqref{erf_asym}
$$
{\rm Erf} (z)=\frac{2z}{\sqrt\pi}e^{-z^2}\bigl[1+O\bigl(z^2\bigr)\bigr],
$$
we obtain for sufficiently long times\footnote{Specifically for $t\gg \omega_0^{-1}$ and $t\gg m_y^2(t)/k^2$. Note that by virtue of Eqs. \eqref{A_0} and \eqref{average_osc_0} $m_y^2(t)/k^2=O(y_0^2/k^2)$.}
\begin{equation}
\mu_u(t)\simeq\frac{\omega_0}{\pi}\exp\left\{-[\omega_0^2(u-m_x(t))^2+m_y^2(t)]/k^2t\right\}.
\label{mu_assym_0}
\end{equation}
Let us finally point out that, although the undamped noisy oscillator is not stationary, its crossing intensity tends as $t\to\infty$ to a finite value independent of any finite crossing level $u$,
\begin{equation}
\lim_{t\to\infty}\mu_u(t)=\frac{\omega_0}{\pi},
\label{mu_limit_0}
\end{equation}
a crossing frequency which doubles the natural frequency of the deterministic oscillator.

\FloatBarrier\subsection{Simulation results}

\begin{figure}[ht]
\includegraphics[width=0.6\linewidth,angle=0]{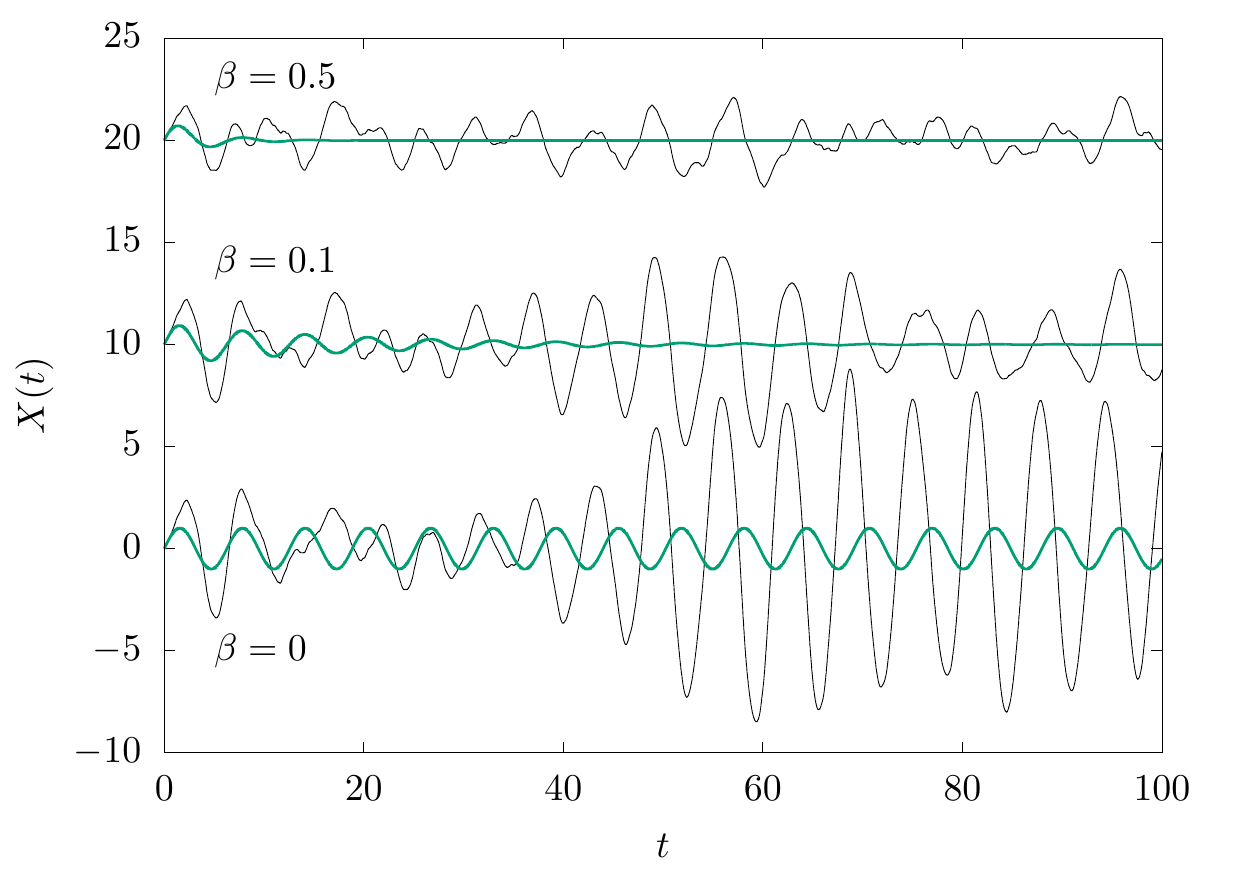}
\caption{Example of random trajectories $X(t)$ for the noisy oscillators with $k=1, \omega_0=1, x_0=0, y_0=1$, shown by the thin (black) lines.
From top to bottom,  $\beta=0.5$, $\beta=0.1$ and $\beta=0$. Data for $\beta=0.5$ and $\beta=0.1$ have been shifted upwards by 20 and 10,
respectively, for better viewing.  Simulations are performed with a fixed time step $dt=0.01$. The thick (green) lines represent the average value $\langle X(t)\rangle$ given in Eq. \eqref{average_osc}.}
\label{traj_noisy_oscillator}
\end{figure}

We have simulated Eq. \eqref{sde_osc} for $\beta \neq 0$ and $\beta=0$ using the algorithm described in Appendix \ref{appendix}. Examples of random trajectories for different values of $\beta$ are shown in Fig. \ref{traj_noisy_oscillator}, together with the average $m_x(t)=\langle X(t)\rangle$.

\begin{figure}[ht]
\includegraphics[width=0.6\linewidth,angle=0]{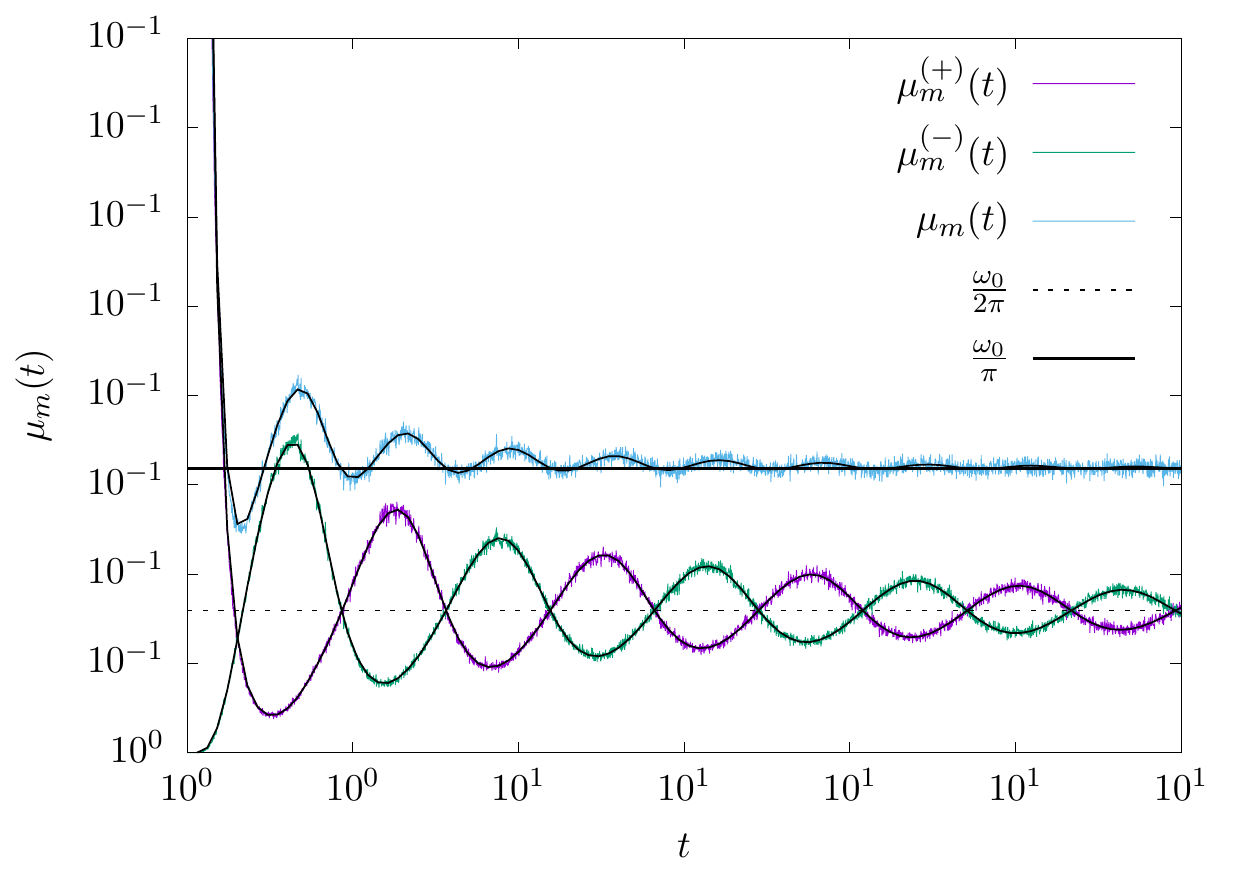}
\caption{The mean-crossing intensity for the damped oscillator with $\beta = 0.1, k=1, \omega_0=1, x_0=0, y_0=1, dt=0.01$.
The noisy colored lines show the simulations results for upcrossing, downcrossing, and total crossing intensities, obtained by averaging over $10^6$ trajectories. The smooth (black) curves are the analytical results, obtained substituting Eqs. \eqref{average_osc}, \eqref{sigma_x_osc}, \eqref{sigma_y_osc}, \eqref{sigma_xy_osc} into Eqs. \eqref{mu+_gauss}, \eqref{mu-_gauss}, \eqref{mu_gauss} for upcrossing, downcrossing, and total intensities, respectively. The horizontal (black) lines show the asymptotic limit for large times.}
\label{mu0_damped_oscillator}
\end{figure}

\begin{figure}[ht]
\includegraphics[width=0.6\linewidth,angle=0]{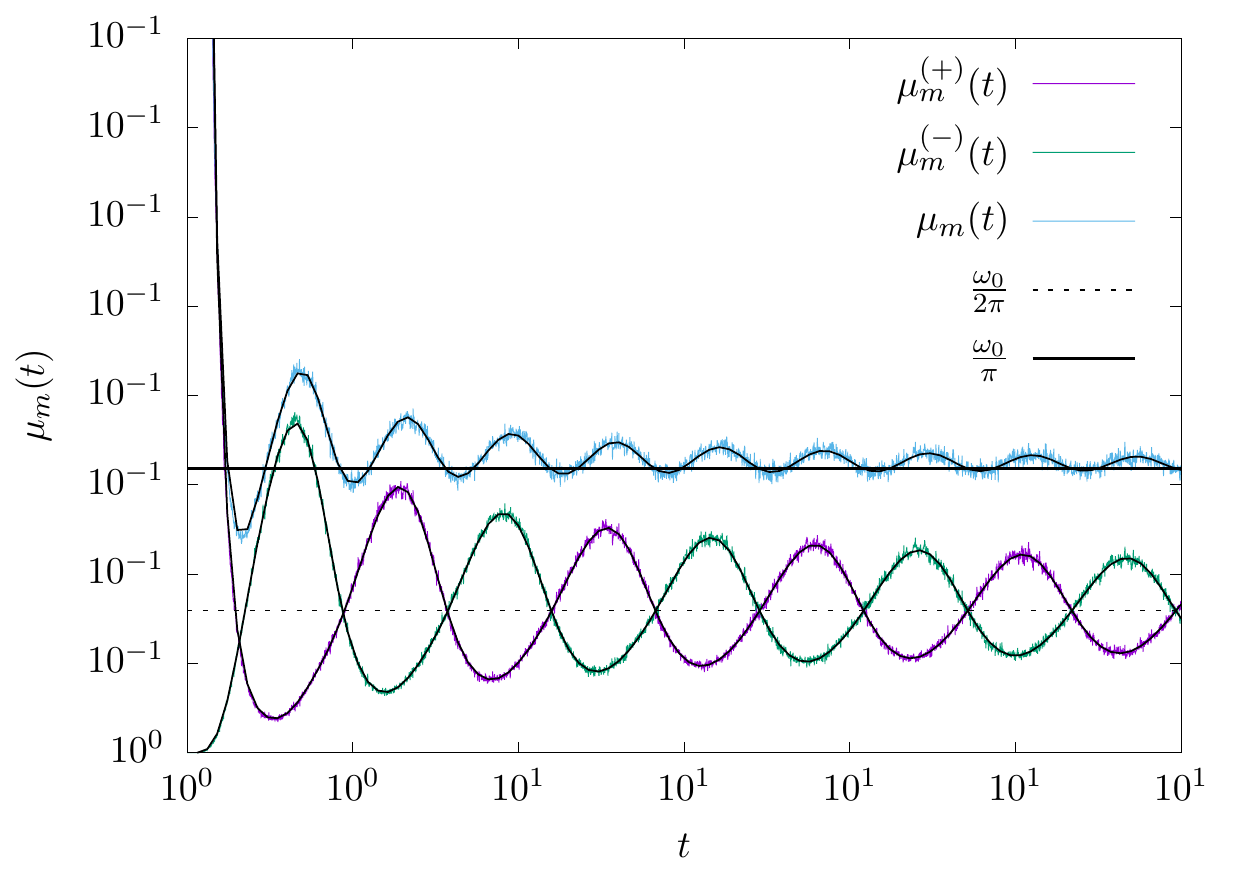}
\caption{The mean-crossing intensity for the undamped oscillator ($\beta=0$) with $k=1, \omega_0=1, x_0=0, y_0=1, dt=0.01$.
The noisy colored lines show the simulations results for upcrossing, downcrossing, and total crossing intensities, obtained by averaging over $10^6$ trajectories. The smooth (black) curves are the analytical results, obtained substituting Eqs. \eqref{average_osc_0}, \eqref{sigma_undamped}, \eqref{delta_osc_0} into Eqs. \eqref{mu+_gauss}, \eqref{mu-_gauss},\eqref{mu_gauss} for upcrossing, downcrossing, and total intensities, respectively. The horizontal (black) lines show the asymptotic limit for large times.}
\label{mu0_undamped_oscillator}
\end{figure}

\begin{figure}[ht]
\includegraphics[width=0.6\linewidth,angle=0]{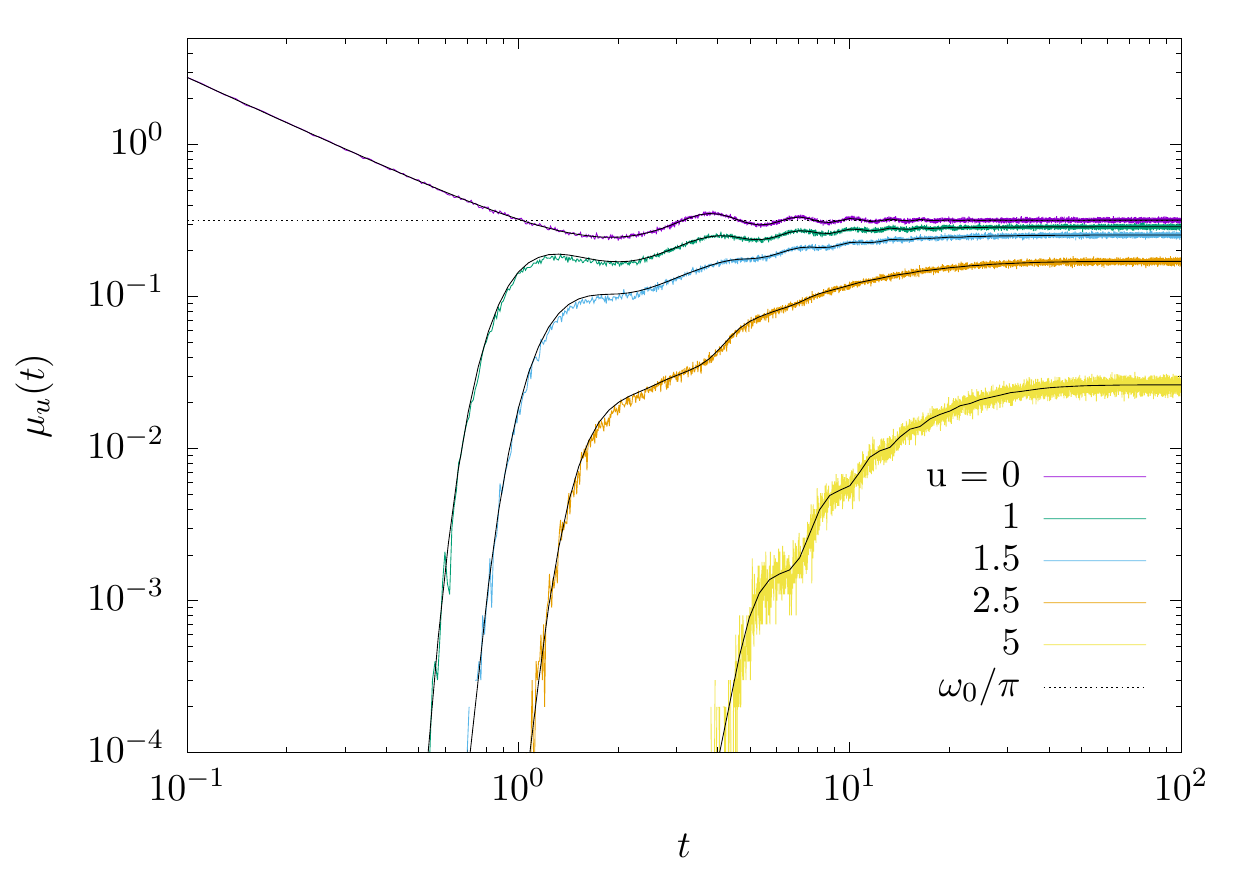}
\caption{Total crossing intensity $\mu_u(t)$ for the damped oscillator with $\beta = 0.1, k=1, \omega_0=1, x_0=0, y_0=0, dt=0.01$, for different levels $u$. The noisy colored lines show the simulations results averaged over $10^6$ trajectories (labels for different $u$ are, from top to bottom, in the same order as the lines). The smooth (black) curves are the analytical results, obtained substituting 
Eqs. \eqref{average_osc}, \eqref{sigma_x_osc}, \eqref{sigma_y_osc}, \eqref{sigma_xy_osc} into Eq. \eqref{mu_gauss}.}
\label{mu_damped_oscillator_y0}
\end{figure}

\begin{figure}[ht]
\includegraphics[width=0.6\linewidth,angle=0]{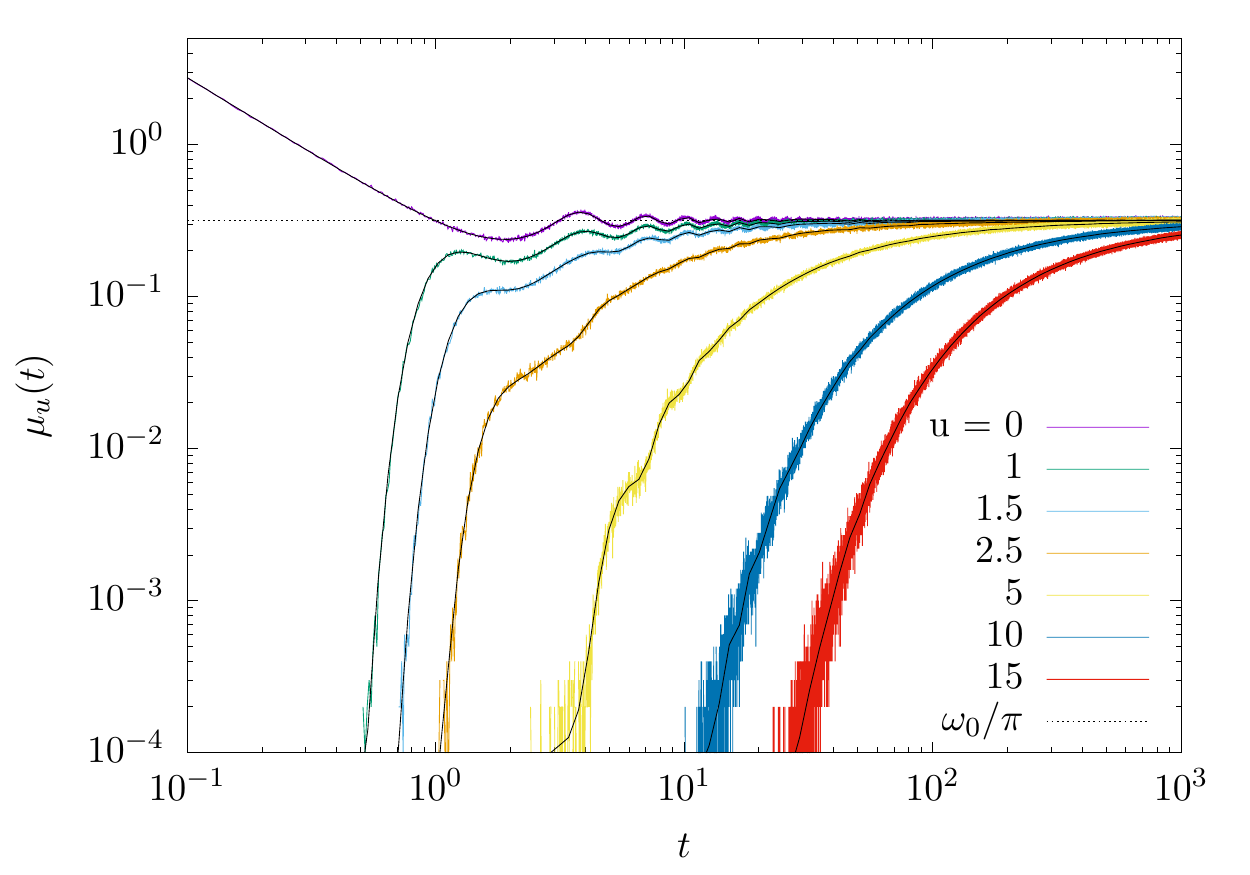}
\caption{Total crossing intensity $\mu_u(t)$ for the undamped oscillator ($\beta=0$) with $k=1, \omega_0=1, x_0=0, y_0=0, dt=0.01$, for different levels $u$.
The colored lines show the simulations results averaged over $10^6$ trajectories (labels for different $u$ are, from top to bottom, in the same order as the lines). The smooth (black) curves are the analytical results, obtained substituting 
Eqs. \eqref{average_osc_0}, \eqref{sigma_undamped}, \eqref{delta_osc_0} into Eq. \eqref{mu_gauss}.}
\label{mu_undamped_oscillator_y0}
\end{figure}

\begin{figure}[ht]
\includegraphics[width=0.6\linewidth,angle=0]{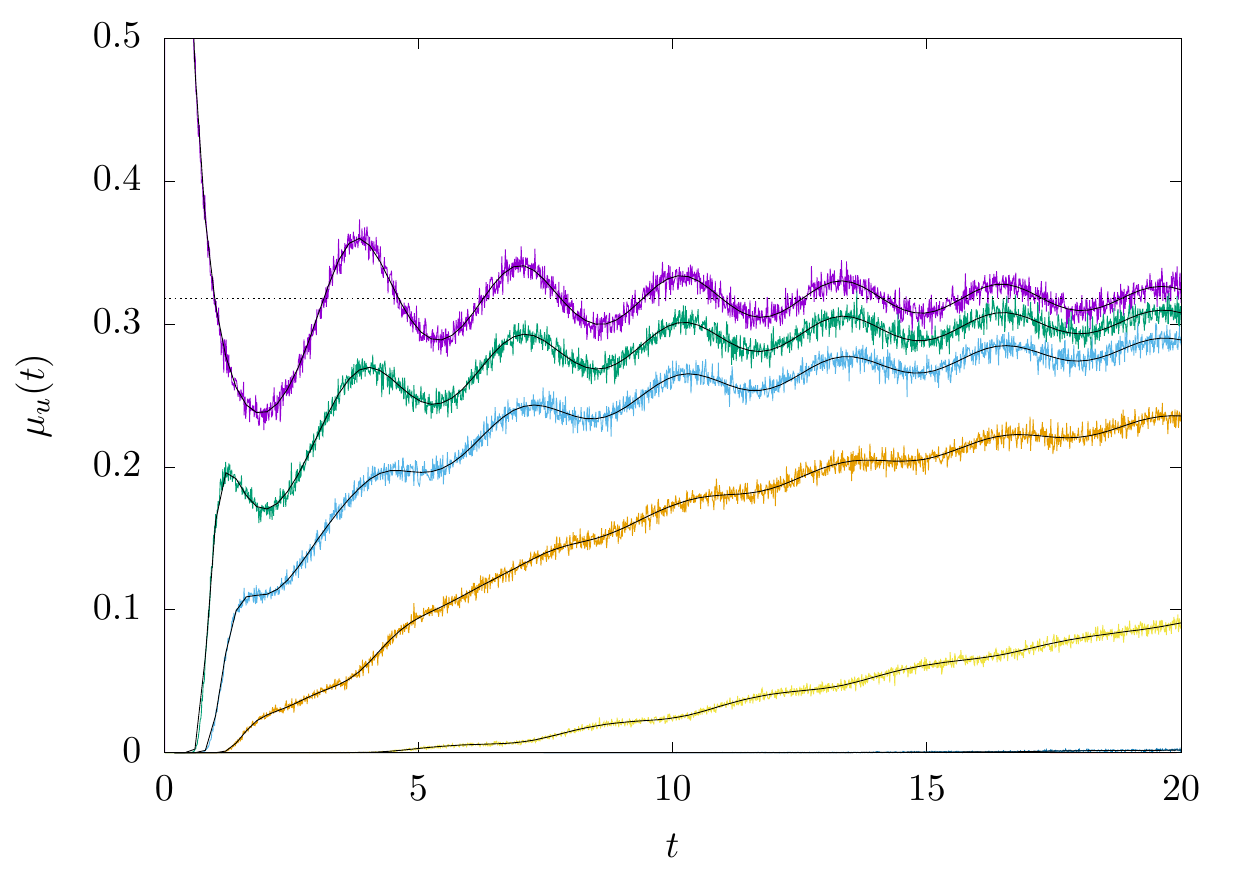}
\caption{Same as Fig. \ref{mu_undamped_oscillator_y0} but in linear scale and short times.}
\label{mu_undamped_oscillator_y0_nolog}
\end{figure}

The mean crossing intensities $\mu_m^{(\pm)}(t)$ and $\mu_m(t)$ as a function of $t$ are shown in Figs. \ref{mu0_damped_oscillator} ($\beta=0$) and \ref{mu0_undamped_oscillator} ($\beta\neq 0$). Note that at large times 
$\mu_m^{(\pm)}(t)$ tend to $\omega_0/(2 \pi)$ and $\mu_m(t)$ tends to $\omega_0/\pi$. In these figures, as well as in the subsequent ones, the smooth black lines show the analytical results which, as they should, are in all cases in perfect agreement with the simulation results.

In comparison with the Brownian motion and random acceleration cases, the noisy oscillator presents an additional time scale $\omega^{-1}$ (or $\omega_0^{-1}$ in the undamped case). When this is much larger than the scales $\beta^{-1}$ and $(y_0/k)^2$, the short-time behavior of $\mu_m(t)$ is the same as that of the Brownian motion.
In particular, if $y_0=0$, then $\mu_m(t) \sim \sqrt{3}/(2 \pi t)$ as $(\beta t,\omega t) \ll 1$, and if $y_0 \neq 0$ then 
$\mu_m \sim (y_0\sqrt 3/2\pi k)t^{-3/2}$ as $t\to 0$. These limits are well verified in the numerical simulations, and scaling plots similar to those for the Brownian motion and random acceleration are obtained (although we do not show them here).

For applications, it is more interesting to examine the behavior of the crossing intensity at a fixed level $u$. 
This is shown in Figs. \ref{mu_damped_oscillator_y0} and \ref{mu_undamped_oscillator_y0} for $\beta\neq 0$ and $\beta =0$, respectively. In both cases we choose zero initial velocity $y_0=0$, so that the symmetry $\mu_u(t)=\mu_{-u}(t)$ holds. 
Note that in the undamped case $\mu_u(t)$ becomes independent of $u$  at large enough times, as predicted analytically.  A detailed view of the behavior at short times for $\beta=0$ is shown in Fig. \ref{mu_undamped_oscillator_y0_nolog} in linear scale.

Analogous plots for $y_0\neq 0$ show a qualitatively similar behavior, except that the symmetry in $u$ is lost at short times.

\FloatBarrier
\section{Concluding remarks}
\label{sec_6}

We have analyzed the counting of crossing events to some preassigned level carried out by inertial random processes. The models studied are described by linear stochastic differential equations of second order driven by Gaussian white noise. The linearity of the equations of motion along with the Gaussian character of the input noise ensure that output processes are Gaussian as well. 

We have firstly reviewed Rice formula for the crossing intensity and generalized it to embrace the most comprehensive kind of Gaussian process. The crossing intensity is an important quantity in many applications. In particular, as we discussed in Section II, its inverse is the return period, which in turn provides an upper bound on the distribution of the maximum of a stochastic process over a given time interval. 
One key result is the exact expression \eqref{mu_gauss} of  the crossing intensity for Gaussian processes in their most general form and the simpler version \eqref{mu_0_b} for the zero crossing, that is, the crossing of the mean value:
$$
\mu_{m}(t)=\frac{ \sigma_y(t)}{\pi \sigma_x(t)} \sqrt{1-\left[\sigma_{xy}(t)/\sigma_x(t)\sigma_y(t)\right]^2}.
$$

We have next specialized on some particular cases of physical interest whose dynamical evolution is described by linear stochastic equations of second oder. In all cases studied we have been able to obtain the exact form for the intensity of up, down and total crossings. 

The simplest example is provided by the random acceleration process, a non-stationary process for which the crossing intensity is time dependent.  
At long times the crossing intensity to any level $u$ decreases with time as
$$
\mu_u(t)\sim t^{-1}, \qquad\quad (t\to\infty),
$$
and thus the average number of crossings increases as
$$
\langle N_u(t)\rangle\sim \ln t, \qquad\quad (t\to\infty).
$$
At short times we find
$$
\mu_u(t)\sim t^{-3/2}, \qquad \langle N_u(t)\rangle\sim t^{-1/2}, \qquad \quad (t\to 0).
$$

The next example is Brownian motion, which is also not stationary.
In this case, at long times (i.e. in the diffusive regime) we obtain a slower decay than that of random acceleration:
$$
\mu_u(t)\sim t^{-1/2}, \qquad \langle N_u(t)\rangle\sim t^{1/2}, \qquad\quad (t\to\infty)\,.
$$ 
For short times (i.e. in the ballistic regime), if the initial velocity is zero ($y_0=0$) we have 
\bd
\mu_u(t)\sim t^{-1}, \qquad \langle N_u(t)\rangle\sim \ln t,  \qquad\quad (t\to 0) \,,
\ed
which is the same scaling as in random acceleration at long times.

The most general case is Brownian motion with non-zero initial velocity ($y_0\neq 0$). This has a more complex time structure since there are now two characteristic time scales. When these scales are well separated we observe three regimes:
$\mu_u(t)\sim t^{-3/2}$ at short times (random acceleration regime), $t^{-1}$ at intermediate times (ballistic regime), and $t^{-1/2}$ at long times (diffusive regime).

The third process studied has been the damped linear oscillator driven by Gaussian white noise. Due to damping, the oscillator reaches a stationary state as time increases, which implies a time-independent crossing intensity that for the mean-crossing problem has the simple expression:
$$
\mu_m=\frac{\omega_0}{\pi},
$$
which doubles the natural frequency of the deterministic oscillator. Let us note that in the stationary state, when transient effects have faded away, the average number of mean  crossings during a time interval $\Delta t$ follows the linear law:
$$
\langle N_m(\Delta t)\rangle=\frac{\omega_0}{\pi} \Delta t.
$$

The last example addressed has been the undamped oscillator. This case is not stationary and the crossing intensity depends on time but tends to a finite and non-zero value as $t\to\infty$ that is independent of the crossing level:
$$
\lim_{t\to\infty} \mu_u(t)=\frac{\omega_0}{\pi},
$$
which again doubles the frequency of the deterministic oscillator. 

Let us finally remark that Rice's approach can be extended to include random processes (whether inertial or not) driven by colored noise as well as to study the counting of maxima and minima. These works are under present investigation and some results will be presented soon.

\appendix

\section{Variances of the Brownian motion}
\label{app1}

From Eqs. \eqref{delta_cor}--\eqref{average_bm} we have
\begin{eqnarray*}
\sigma_x^2(t)&=&\left\langle\left[X(t)-m_x(t)\right]^2\right\rangle\\
&=&\frac{k^2}{\beta^2}\int_0^t dt_1\int_0^t\left[1-e^{-\beta(t-t_1)}\right]\left[1-e^{-\beta(t-t_2)}\right]\delta(t_1-t_2)dt_2 \\
&=& \frac{k^2}{\beta^2}\int_0^t \left[1-e^{-\beta(t-t_1)}\right]^2 dt_1,
\end{eqnarray*}
and hence
\begin{equation*}
\sigma_x^2(t)=\frac{k^2}{\beta^3}\left(\beta t-\frac 32 + 2e^{-\beta t}-\frac 12 e^{-2\beta t}\right),
\end{equation*}
which agrees with Eq. \eqref{sigma_x_bm}. Proceeding in an analogous way, we have
\begin{eqnarray*}
\sigma_y^2(t)&=&\left\langle\left[Y(t)-m_y(t)\right]^2\right\rangle\\
&=&k^2\int_0^t dt_1\int_0^t e^{-\beta(t-t_1)} e^{-\beta(t-t_2)}\delta(t_1-t_2)dt_2\\
&=&k^2 \int_0^t e^{-2\beta(t-t_1)} dt_1,
\end{eqnarray*}
and we obtain Eq. \eqref{sigma_y_bm}:
\begin{equation*}
\sigma_y^2(t)=\frac{k^2}{2\beta}\left(1-e^{-2\beta t}\right).
\end{equation*}
Finally, 
\begin{eqnarray*}
\sigma_{xy}(t)&=&\Bigl\langle [X(t)-m_x(t)][Y(t)-m_y(t)]\Bigr\rangle \\
&=&\frac{k^2}{\beta}\int_0^t dt_1\int_0^t e^{-\beta(t-t_1)}\left[1-e^{-\beta(t-t_2)}\right]\delta(t_1-t_2)dt_2 \\
&=&\frac{k^2}{\beta}\int_0^t e^{-\beta(t-t_1)}\left[1-e^{-\beta(t-t_1)}\right] dt_1,
\end{eqnarray*}
that is, 
\begin{equation*}
\sigma_{xy}(t)=\frac{k^2}{\beta^2}\left(\frac 12 - e^{-\beta t}+\frac 12 e^{-2\beta t}\right).
\end{equation*}
which is Eq. \eqref{sigma_xy_bm}.

\section{Simulation method}
\label{appendix}

We will use the algorithm presented in Ref.\cite{farago} to simulate the Langevin equation  
\bd
\ddot{X}+\beta \dot{X} = F(X,t) + k\xi(t),
\ed
where $\xi(t)$ is Gaussian white noise satisfying $\langle \xi(t) \xi(t^\prime)\ra = \delta(t-t^\prime)$ and 
$F(X(t),t)$ is a deterministic force. 

Discretizing time as $t_{n+1}=t_n + dt$, a random trajectory
$(t_n, X_n)$, starting from the initial condition $X_0 = x_0$, $Y_0= y_0$, where $Y=\dot{X}$, is generated
by iterating the following recursive equations (in our notation):
\beqn
X_{n+1}&=&X_n + b dt Y_n + \frac{b}{2}(dt)^2 F_n  + \frac{b}{2} k (dt)^{3/2} g_{n+1} \\
Y_{n+1}&=&Y_n + \frac{b}{2} dt (F_n+F_{n+1}) - \beta (X_{n+1}-X_n) + k (dt)^{1/2} g_{n+1}
\eeqn
where $F_n=F(X_n,t_n)$, $b=(1+\beta/2)^{-1}$, and the $g_n$ are i.i.d. Gaussian random variables
with $\langle g_n\rangle=0, \langle g_n^2\rangle=1$.

Averaging over $R$ trajectories, we measure the up- and down-crossing intensities $\mu_u^+(t_n),\mu_u^-(t_n)$ 
at each $t_n$, where for example $\mu_u^+$ is total number of upcrossings taking place in $(t_n,t_{n+1})$
(we say an upcrossing has taken place if $X_n < u$ and $X_{n+1}>u$), divided by $R \,t_n$.
The total crossing intensity is $\mu_u(t_n)=\mu_u^+(t_n)+\mu_u^-(t_n)$.

In our numerical results, $X$ and $t$ are in arbitrary units. It is helpful to think of $X$ as a length expressed in meters, and $t$ expressed in seconds. Then, the units of the parameters are as follows: $[\beta] =$s$^{-1}$, $[k]=$m s$^{-3/2}$, 
$[y_0]=$m s$^{-1}$, $[\omega_0, \omega]=$s$^{-1}$.
For Brownian motion and random acceleration, we typically use a time step $dt = \alpha \sqrt{t}$ with $\alpha=10^{-3}$ for $t<1$ and $\alpha = 10^{-2}$ for $t>1$, except for large $\beta$ for which we choose $\alpha$ to be ten times smaller. 

In all cases, we set $R=10^6$.

\section{Variances of the noisy oscillator}
\label{app2}

From Eqs. \eqref{osci_x}, \eqref{osci_y} and \eqref{average_osc}, we have
\begin{eqnarray*}
\sigma_x^2(t)&=&\left\langle\left[X(t)-m_x(t)\right]^2\right\rangle\\
&=&\frac{k^2}{\omega^2} e^{-\beta t}\int_{t_0}^t dt_1\int_{t_0}^t e^{\beta(t_1+t_2)/2} \sin\omega(t-t_1) \sin\omega(t-t_2)\delta(t_1-t_2) dt_2\\
&=&\frac{k^2}{\omega^2} e^{-\beta t}\int_{t_0}^t e^{\beta t_1}\sin^2\omega(t-t_1)dt_1=\frac{k^2}{\omega^2} \int_{0}^{t-t_0} e^{-\beta t'}\sin^2 \omega t'dt',
\end{eqnarray*}
hence
\begin{equation*}
\sigma_x^2(t)=\frac{k^2}{\omega^2(\beta^2+4\omega^2)}\left\{\frac{2\omega^2}{\beta}-
e^{-\beta(t-t_0)}\left[\beta\sin^2\omega(t-t_0)+\omega\sin 2\omega(t-t_0)+\frac{2\omega^2}{\beta}\right]\right\},
\end{equation*}
which is Eq. \eqref{sigma_x_osc}.
Proceeding in an analogous way, we have
\begin{eqnarray*}
\sigma_y^2(t)&=&\left\langle\left[Y(t)-m_y(t)\right]^2\right\rangle\\
&=&k^2 e^{-\beta t}\int_{t_0}^t dt_1\int_{t_0}^t e^{\beta(t_1+t_2)/2} \cos\omega(t-t_1) \cos\omega(t-t_2)\delta(t_1-t_2) dt_2\\
&=&k^2 \int_{0}^{t-t_0} e^{-\beta t'}\cos^2 \omega t'dt',
\end{eqnarray*}
and
\begin{equation*}
\sigma_y^2(t)=\frac{k^2}{\beta^2+4\omega^2}\left\{\frac 1\beta(\beta^2+2\omega^2)-
e^{-\beta(t-t_0)}\left[\beta\cos^2\omega(t-t_0)-\omega\sin 2\omega(t-t_0)+\frac{2\omega^2}{\beta}\right]\right\},
\end{equation*}
which agrees with Eq. \eqref{sigma_y_osc}. Finally,
\begin{eqnarray*}
\sigma_{xy}(t)&=&\Bigl\langle [X(t)-m_x(t)][Y(t)-m_y(t)]\Bigr\rangle \\
&=& \frac{k^2}{\omega} e^{-\beta t}\int_{t_0}^t dt_1\int_{t_0}^t e^{\beta(t_1+t_2)/2} \sin\omega(t-t_1) \cos\omega(t-t_2)\delta(t_1-t_2) dt_2 \\
&=& \frac{k^2}{2\omega} e^{-\beta t}\int_{t_0}^t e^{\beta t_1} \sin 2\omega(t-t_1)dt_1=
\frac{k^2}{2\omega} \int_0^{t-t_0} e^{-\beta t'} \sin 2\omega t' dt',
\end{eqnarray*}
that is, 
\begin{equation*}
\sigma_{xy}(t)=\frac{k^2}{2\omega(\beta^2+4\omega^2)}\left\{2\omega-e^{-\beta(t-t_0)} 
\left[\beta\sin 2\omega(t-t_0)+2\omega\cos 2\omega(t-t_0)\right]\right\},
\end{equation*}
which is Eq. \eqref{sigma_xy_osc}.

For the undamped oscillator $\beta=0$ and from the above equations we have (recall we have set $t_0=0$)
$$
\sigma_x^2(t)=\frac{k^2}{\omega_0^2} \int_{0}^{t} \sin^2 \omega_0 t'dt'=\frac{k^2t}{2\omega_0^2}\left(1-\frac{1}{2\omega_0t} \sin 2\omega_0 t\right),
$$
$$
\sigma_y^2(t)=k^2 \int_{0}^{t} \cos^2 \omega_0 t'dt'=\frac{k^2t}{2}\left(1+\frac{1}{2\omega_0t} \sin 2\omega_0 t\right),
$$
$$
\sigma_{xy}(t)= \frac{k^2}{2\omega_0}\int_{0}^{t} \sin 2\omega_0 t' dt'=\frac{k^2}{4\omega_0^2}\left(1-\cos 2\omega_0 t\right),
$$
which agree with Eq. \eqref{sigma_undamped}.

\acknowledgments 
This work has been partially funded by MINECO (Spain), Agencia Estatal de Investigaci\'on (AEI) grant numbers PID2019-106811GB-C33 (AEI/10.13039/501100011033) 
(J.M.), PGC2018-094754-B-C22 (M.P.), and by Generalitat de Catalunya grant numbers 2017SGR608 (J.M.), 2017SGR1614 (M.P.). M.P. thanks Marco Palassini Vidal for inspiration.

\end{document}